%% file: ms.tex
\documentclass[12pt,preprint]{aastex}
\shorttitle{M101 X-ray Sources}
\shortauthors{Pence et al.}

\begin{document}


\title{{\em Chandra} X-ray Sources in M101}


\author{W. D. Pence, S. L. Snowden\altaffilmark{1}, 
and K. Mukai\altaffilmark{1}}
\affil{Code 662, NASA/Goddard Space Flight Center, Greenbelt, MD 20771}

\and

\author{K. D. Kuntz}
\affil{Joint Center for Astrophysics, Department of Physics, University of Maryland,
Baltimore County, 1000 Hilltop Circle, Baltimore MD 21250}
%\email{aastex-help@aas.org}


\altaffiltext{1}{Universities Space Research Association}


\begin{abstract}

A deep (98.2 ks) {\em Chandra} Cycle-1 observation has revealed a
wealth of discrete X-ray sources as well as diffuse emission in the
nearby face-on spiral galaxy M101.  From this rich dataset we have
created a catalog of the 110 sources from the S3 chip detected with a
significance of $>3 \sigma$.  This detection threshold corresponds to a
flux of $\sim10^{-16}$ ergs cm$^{-2}$ s$^{-1}$ and a luminosity of
$\sim10^{36}$ ergs s$^{-1}$ for a distance to M101 of 7.2 Mpc.  The
sources display a distinct correlation with the spiral arms of M101 and
include a variety of X-ray binaries, supersoft sources, supernova
remnants, and other objects of which only $\sim27$ are likely to be
background sources.  There are only a few sources in the interarm
regions, and most of these have X-ray colors consistent with that of
background AGNs.  The derived $\log N-\log S$ relation for the sources
in M101 (background subtracted) has a slope of $-0.80 \pm 0.05$ over
the range of $10^{36} - 10^{38}$ ergs s$^{-1}$.  The nucleus is
resolved into 2 nearly identical X-ray sources, each with a 0.5 -- 2.0
keV flux of $4 \times 10^{37}$ ergs s$^{-1}$.  One of these sources
coincides with the optical nucleus, and the other coincides with a
cluster of stars 110 pc to the south.  The field includes 54 optically
identified SNR, of which 12 are detected by {\em Chandra}. Two of the
SNR sources are variable and hence must be compact objects.  In total,
8 of the X-ray sources show evidence for short term temporal variation
during this observation.  Two of these variable sources are now
brighter than the {\em ROSAT} detection threshold, but they were not
detected in the previous {\em ROSAT} observations taken in 1992 and
1996.  There are also 2 variable sources previously seen with {\em
ROSAT} that apparently have faded below the {\em Chandra} detection
threshold.  The brightest source in the field shows extreme long-term
and short-term temporal variability.  At it's peak brightness it has a
super-Eddington  luminosity $>10^{39}$ ergs s$^{-1}$.  There are 10
Supersoft sources (SSS) in the field which can be divided into 2
distinct subclasses:  the brighter class (3 objects) has a luminosity
of $\sim10^{38}$ ergs s$^{-1}$ and a blackbody temperature of $\sim70$
eV whereas the other class (7 objects) is an order of magnitude fainter
and has a blackbody temperature of only $\sim50$ eV.

\end{abstract}




\section{Introduction}

M101 is a nearby (7.2 Mpc, \citet{ste98}), face-on spiral galaxy that
is ideal for studies of the X-ray source population in a galaxy similar
to the Milky Way.  Since M101 is viewed through a relatively small
amount of obscuring foreground Galactic hydrogen gas ($N_H = 1.2 \times
10^{20}$  cm$^{-2}$, \citet{sta92}), it is especially well suited for
surveying soft X-ray sources, without the observational biases
introduced by the thick absorption in the Galactic plane when studying
X-ray sources in our own Galaxy.  Although M101 has a slightly later
morphological type (Scd vs. Sbc) and a significantly larger D$_{25}$
isophotal diameter (60 kpc vs. 23 kpc) than the Milky Way
\citep{dev78,dev91}, one would expect that the properties of the X-ray
source populations in these 2 galaxies should be quite similar.

M101 was previously observed with the {\em Einstein} X-ray observatory
\citep{mcc84, tri90}, and later with {\em ROSAT} to study both the
discrete sources \citep{wan99} and the diffuse X-ray emission
\citep{sno95}.   M101 was an obvious candidate for follow up studies
with the {\em Chandra} X-ray observatory, and this paper presents the
basic data and analysis of the population of discrete X-ray sources
detected in a Cycle-1 {\em Chandra} observation.  This paper is one in
a series which analyze various aspects of this rich dataset.  A
previous paper \citep{sno01} reinvestigated some of the hypernova
remnant candidates in M101, and 2 other papers currently in
preparation will study the brighter black hole X-ray binary
candidates in more detail and analyze the diffuse X-ray emission in
M101.

\section{Data Processing}

The present data set consists of a 98.2 ks exposure taken on 26--27
March 2000 with the {\em Chandra} observatory ACIS imaging spectrometer
with the nucleus of M101 centered near the ``aim point'' of the S3 CCD
chip.  The analysis in this paper is limited to the sources detected
within the 8\farcm4 x 8\farcm4 (17.6 x 17.6 kpc) field of view of the
S3 chip which has much better angular resolution and extended
sensitivity to lower X-ray energies than the other chips.  This chip
covers the central region of M101 out to a minimum distance of 6.3 kpc
from the nucleus towards the NW, and a maximum distance of 14.9 kpc at
the east corner of the chip.  By comparison, the prominent spiral arms
extend out to about 14 kpc, and the $R_{25}$ isophotal radius
\citep{dev91} is 30.2 kpc, so the S3 chip covers about 50\% and 11\%,
respectively, of these circular areas in M101.

The standard CIAO pipeline software system (``Rev 2'' as of December
2000) was used to clean and calibrate the data set.   In addition to
the standard data cleaning, we also excluded the data during a 64 s
interval of unusually high background count rate which reduced the
exposure duration to 98180 s.  Finally, we excluded all photon events
which had a nominal pulse invariant (PI) channel energy greater than
8.0 keV where the S3 chip has virtually no sensitivity.  The nominal
lower energy cutoff in the data set is 0.125 keV.

\section{Source Detection}  \label{detections}

The CIAO {\em celldetect}\footnote{
At the time that we received the Chandra data files the {\em
celldetect} program appeared to produce the most reliable list of
sources.  Later, after the release CIAO v2.1, we repeated the source
detection procedure using the {\em wavdetect} program, but concluded
that the final list of sources presented in Table 1 would not change
significantly.  The $\log N-\log S$  analysis in \S\ref{lognlogs} was
based on the results from {\em wavdetect}.}
program was initially used with a threshold of $2.5
\sigma$ to generate a preliminary candidate list of X-ray sources in
the 0.125 -- 8.0 keV band image. At this low threshold setting about
20\% of the sources were clearly spurious, mainly located in the wings
of brighter sources near the edges of the chip.  These spurious sources
were rejected by inspecting their positions on an adaptively smoothed
image of the S3 chip field of view constructed from all the X-ray
events in which sources with as few as 10 net counts are clearly
visible.  The final adopted set of sources have formal signal to noise
ratios close to 3.0 or higher.  Note that the detection threshold for
faint point sources increases with distance from the ``aim point''
because of the decreased angular resolution, and as a result the
faintest point sources (with $F_x \sim10^{-16}$ ergs cm$^{-2}$
s$^{-1}$) are detectable only within a radius of about 2\arcmin ~= 4
kpc from the nucleus of M101.

The positions of our final 110 {\em Chandra} X-ray sources in M101 are
shown in Figures 1 and 2, and listed in Table 1.  The source positions
derived from the {\em Chandra} data are generally accurate to better
than 1\arcsec.  The mean difference between our X-ray positions and the
astrometric optical positions of 12 previously identified supernovae
remnants catalogued by \citet{mat97} is only 0\farcs47 $\pm$ 0\farcs20,
about the size of one ACIS CCD pixel.

As can be seen in Figure 2, the X-ray sources are strongly correlated
with the position of the optical spiral arms and bright HII regions in
M101.  There are relatively few X-ray sources in the interarm regions
which indicates that most of the X-ray sources are associated with the
young stellar population in M101.  To quantify this association, we
divided the area covered by the S3 chip into  either ``spiral arm'' or
``interarm'' regions based on the appearance in the blue optical
photograph of M101, with the result that 75\% of the area of the chip
was classified as covering spiral arm regions (also includes the
nuclear region) and 25\% as covering interarm regions.  Using this
spatial criteria, only 8 of the  X-ray sources (sources 5, 7, 8, 19,
29, 39, 64, and 97) were found to be located in the interarm regions
and not associated with optical emission in M101.  This gives a mean
surface density of 1.9 sources arcmin$^{-2}$ in the spiral arms
compared with only 0.45 sources arcmin$^{-2}$ in the interarm regions
(both before background subtraction).

We calculated the expected number of background sources in our field
based on the point source luminosity function determined from a similar
{\em Chandra} observation of a blank field by \citet{mus00}, the HI
column density map of M101 by \citet{kam93}, and the \citet{har97}
measure of the Galactic HI column.  Given that the {\it Chandra}
response is rather hard and that the AGN spectra are also hard, the
diminution of the surface density of AGNs due to the HI absorption in
M101 is rather small, only about 8\%.  From this we estimate that about
27 of the detected sources, only a quarter of our total sample, are
likely to be serendipitous background (or perhaps in a few cases,
foreground) sources.  Based on the above division of the chip area into
``spiral arm'' and ``interarm'' regions, about 20 of these background
sources should appear to be located in the spiral arm and nuclear
regions.  This corresponds to a mean background surface density of 0.38
sources arcmin$^{-2}$ which is remarkably close to the measured density
of sources in the ``interarm'' regions of 0.45 sources arcmin$^{-2}$.
This suggests that most of these interarm sources are background AGNs,
and indeed, 6 of the 8 interarm sources do have relatively hard SR1 and
HR1 colors (see \S\ref{colors} and Fig. 6) which are consistent with
the expected power law spectrum of an AGN.  Furthermore, the brightest
of these sources (\#5) was previously identified as an AGN by
\citet{wan99}.  The other 2 sources located in the interarm regions
(\#7 and \#8) have very soft spectra and thus are more likely to be
physically associated with M101.  Source \#8 in fact is one of the
Supersoft sources discussed in \S\ref{SSS}.

\section{The log N--log S Relation} \label{lognlogs}

To create a $\log N-\log S$ relation for the point sources towards M101
that would be more easily comparable to relations derived for other
galaxies, we applied somewhat more restrictive criteria than for the
catalogue of sources.  We chose all sources where the effective
exposure was greater than 75\% of the maximum and having a significance
greater than 4.5 in the 0.5-2.0 keV energy band (the M2 band defined in
\S\ref{colors}), where the source counts were measured within the
elliptical PSF region containing 75\% of the encircled energy, and the
background was measured from an annular region 2 to 4 times the size of
the 95\% encircled energy ellipse.  The source rates were corrected for
vignetting.  (This choice of energy range balances the need to detect
the most numerous types of objects in M101, which are soft, with the
desire to detect hard objects as well.) The object catalogue was binned
into $\Delta\log(S/\mbox{counts s}^{-1})$ bins, and each bin was
divided by the number of square arcminutes over which such a flux could
have been 
detected.\footnote{The point source detection limit is
spatially variable, as both the size of the PSF and the background rate
vary.  Although the variation of the size of the PSF dominates, since
some point sources are embedded in diffuse emission, the background
rate can be significant in determining the point source detection
limit.  Both effects, as well as the effect of the vignetting, were
included in this analysis.}
The $\log N-\log S$ relation is shown in
Figure~\ref{fig:lnls} along with the relation due to background AGN
seen through the disk of M101 (as calculated from the relation of
\citet{mus00}).  A flux conversion factor equivalent to $10^{-14}$ ergs
cm$^{-2}$ s$^{-1}$ = $3.33 \times 10^{-3}$ counts s$^{-1}$, which is
appropriate for a background AGN with a $\Gamma=1.42$ power-law X-ray
spectrum, was used in these calculations.  The resulting $\log N-\log
S$ relation for the sources not due to the background AGN is nearly
linear with $N(>S)\propto S^{-0.80\pm0.05}$.  Excluding the 8 interarm
sources identified in the previous section from the analysis only makes
a slight difference, resulting in a slope of $-0.76\pm0.05$. The
uncertainty of the slope was calculated following the \cite{lam76}
criteria, and is larger than the bootstrap uncertainty ($\pm0.02$) or
the variation produced by changing the flux range of the fit
($\pm0.02$).  The relatively flat slope derived here implies that the
amount of diffuse emission due to unresolved faint point sources will be
small.

Figure~\ref{fig:rad} shows the surface density of point sources as a
function of radius from the nucleus.  The optical bulge of M101 is
visible out to $\sim0\farcm75$ \citep{oka76}, and we do indeed see an
increase in the point source surface density in the bulge region.
Because the total number of sources within the bulge region is small
($\sim14$) it is difficult to determine whether the ``bulge sources''
follow a different $\log N-\log S$ relation than the ``disk sources''.
However, to within the uncertainties, we do not observe a difference in
the shapes of the relations between bulge ($R<1\arcmin$) and disk
sources ($1\arcmin<R<4.5\arcmin$).

The above analysis could be repeated for the 2.0--8.0 keV band using
the 2-10 keV band AGN luminosity function found by \citet{mus00}, but
in this energy range only $\sim18$ of the objects detected with
$\sigma>4.5$ are not background AGN. The slope of the $\log N-\log S$
relation is quite uncertain ($0.69\pm0.18$, where the quoted
uncertainty is no more than a formal uncertainty); to within the
uncertainties, the 2.0--8.0 keV band is consistent with the 0.5--2.0
band.

That the slope of our 0.5--2.0 keV luminosity function does not agree
with that of \citet{wan99} ($1.9^{+1.3}_{-0.3}$) is hardly
surprising; the HRI data was over an order of magnitude less sensitive,
many of the sources detected are actually composed of several point
sources, and some of the brighter sources are embedded in diffuse
emission that would have fallen within the HRI PSF.  Further, the Wang
et al. (1999) luminosity function was formed over a much larger area (a
radius of $12\arcmin$ compared to our $\leq6\arcmin$) and thus may be
composed of a different population of sources.

\section{Source Identification}

About 1/4 of the {\em Chandra} X-ray sources can be cross-identified
with previously observed X-ray sources or are positionally coincident
with other catalogued objects as listed in Table 2.  The previous {\em
ROSAT} HRI and PSPC observations of M101 \citep{wan99} detected 19
sources within our field of view, and all but 3 of these are detected
in the {\em Chandra} observation.  A 20th source (their source H42) is
just visible on the edge of the field, but it is too highly distorted
and affected by the edge of the detector to be included in our final
list of sources.  Of the 3 non-detected sources, 2 of them were
classified as variable by \citet{wan99} (their sources H28 and H31) so
they may have been in a low flux state during our observation.  The
other non-detected object (their H41 source) is located near the East
corner of the S3 chip, where our detection threshold is relatively
high.

The nuclear source seen in the  {\em ROSAT} HRI image (source H23), is
resolved into 2 nearly identical point sources in the {\em Chandra}
image.  The northern component (source \#40) appears to coincide with
the nucleus at $14^h03^m12.55^s$, $54^{\circ}20$\arcmin56\farcs5 but is
otherwise unremarkable compared to the other X-ray sources in the
field.  The other nuclear component is 3\farcs1 ($\sim110$ pc) to the
south and coincides with a loose cluster of bright stars as seen in an
HST WFPC-I optical image (after applying a $\sim1$\arcsec ~offset to
the nominal HST coordinates to co-align the nucleus in both images).
The nuclear X-ray source is slightly brighter than the southern
component (they are ranked 9th and 12th in relative X-ray count rate in
Table 1), and has a slightly softer spectrum, although both sources
have fairly neutral X-ray hardness ratios. 

\citet{mat97} identified 54 optical SNRs within the {\em Chandra} field
of view and 9 of these were detected in the X-ray band and listed in
Table 2.  Another 3 SNR sources (their SNR numbers 24, 32, and 43) are
faintly visible on the adaptively smoothed {\em Chandra} image, but the
detections were not significant enough to be included in our final list
of 110 sources.  All 9 of the X-ray detected SNRs sources have soft
X-ray spectra, as would be expected from a hot plasma.  The 2 brightest
SNR sources (our \#85 and \#104) were variable during the {\em Chandra}
observation, so in at least these cases the X-ray flux must be coming
from a compact source and not from from the extended SNR itself.  In a
companion paper \citep{sno01} we present a more detailed investigation
into the properties of several of these SNRs.

To complete the Table 2 list of cross identifications, 2 of the 5 giant
HII regions in M101 \citep{wil95} are located near the edge of the S3
chip (source \#107 is  NGC 5461  and source \#110 is NGC 5462). The
supernova SN1951H was located about $12''$ SE of NGC 5462
\citep{isr75}, but it is not detected in the {\em Chandra}
observation.  Finally, 7 of the {\em Chandra} sources are coincident
with near-IR objects in the 2MASS catalog and 2 of these are also cross
identified with sources in the HST Guide Star Catalog (one being the
nucleus, the other being NGC 5458, a bright HII region).

About 10\% of globular clusters in our Galaxy and other nearby galaxies
contain X-ray emitting neutron stars, so a population of globular
clusters might account for some of the X-ray sources in M101.  The only
published list of globular type clusters in M101 \citep{bre96} contains
41 visually identified clusters within one relatively small (2\arcmin
~square) HST WFPC II observation.  It turns out, however, that most of
these clusters are very blue and hence presumably young, like those
observed in the Magellanic Clouds.  Out of the 14 X-ray sources in this
same region of M101, only one of them (source \#80, one of the
Supersoft sources discussed in \S\ref{SSS}) is located within 2\arcsec
~of any of the optical clusters, which has a $\sim40$\% probability of
occurring by chance.

\section{X-ray Colors} \label{colors}

The X-ray events located within a circular region around each of the
110 X-ray sources were used to calculate the count rate within standard
energy bands and to generate spectra and light curves for each source.
The radius of the region was chosen as a function of the distance of
the source from the optical axis so as to enclose $>95$\% of the source
flux as predicted from the model PSFs and verified by numerical
experiments on the brightest sources.  The area of each source region
is given by the ``Area'' column in Table 1, in units of the 0\farcs49
{\em Chandra} CCD pixels.

Since most of the sources are too faint for detailed spectral analysis,
we have characterized the general spectral properties of the sources by
using the ratios of the X-ray count rates in 2 sets of soft, medium, and
hard energy X-ray bands.  The energy ranges for the first set of bands
(S1, M1, and H1) were selected to produce roughly equal count rates in
each band for the typical sample of sources seen in M101.  For the
second set of bands (S2, M2, and H2) a more standard set of
astrophysically relevant energy ranges were used to facilitate the
comparison of our results with those from previous X-ray missions.  The
adopted energy range for each of these bands is given in Table 3 along
with the observed mean background flux level in each band.  (The slight
variation in background level and diffuse emission across the field is
not significant for these purposes).  The net counts, after background
subtraction, for each source in each of these 6 bands is listed in
Table 1.  A pair of soft and hard X-ray flux ratios (or X-ray colors)
were then calculated from the net counts in the three bands, where  SR1
= (M1 - S1) / (S1 + M1), HR1 = (H1 - M1) / (M1 + H1), and similarly for
the SR2 and HR2 ratios.  In principle, the measured colors will depend
on the position of the source in the field because the sensitivity of
the ACIS instrument (i.e., the effective area of the telescope)
decreases with distance from the aimpoint at slightly different rates
depending on the energy band.  This effect is small over the area
covered by the S1 chip, however, and would cause the apparent X-ray
colors to shift by only about 0.01 as the source is moved from the aim
point to the far corner of the chip.

We compared the {\em Chandra} 0.125-2.0 keV (S2 + M2 band) count rates
with the {\em ROSAT} HRI count rates as measured by \citet{wan99}.
Figure 5 shows that there is a very good correlation between the HRI
and {\em Chandra} count rates for 15 of the 16 sources in common, which
indicates that they have not varied significantly between the epochs of
the observations (80\% of the HRI data were obtained in 1996 and 20\%
in 1992).  The one very discrepant point is source \#98 which is highly
variable (see \S\ref{timevar}) and is now about 25 times brighter in
mean flux level than during the {\em ROSAT} observations.  It has been
plotted at 1/10 of its actual mean{\em Chandra} count rate in order to
fit onto the scale of Figure 5.  The other bright source in Figure 5
(NGC 5462) has a slightly higher {\em ROSAT} count rate than would be
predicted by the relation defined by the other points, but the
comparison is complicated because the higher resolution {\em Chandra}
image shows that it is a blend of several sources.  Also shown at the
bottom of Figure 5 are the {\em Chandra} count rates for 11 sources
which were not detected in the {\em ROSAT} observations.  The brightest
2 of these sources (\#93 and \#71) were variable during the {\em
Chandra} observation (and hence could have been in a low state during
the HRI observations), and the other sources are close to the {\em
ROSAT} detection threshold, so it is not particularly surprising that
these sources were not previously detected the {\em ROSAT} data. 

The SR1 and HR1 X-ray colors provide a sensitive discriminator between
sources with different types of X-ray spectra as shown in Figure 6.
(Note that the SR2 and HR2 colors are less useful for this purpose
because most of the flux lies in the M2 band and hence most of the
points end up crowded into the lower right quadrant of the HR2 vs. SR2
diagram).  The 45 brightest sources, which have more than 40 net X-ray
counts, have been plotted with their calculated $1\sigma$ error bars.
The fainter sources have been omitted for clarity.  Any points which
have a SR1 or HR1 value lying outside the physically allowed range of
-1 to +1 (due to uncertain background subtraction) have been plotted at
the limiting value.  It can be seen that most of the points in Figure 6
lie within a broad diagonal band ranging from the softest sources in
the lower left to the hardest sources in the upper right.

The calculated SR1 and HR1 flux ratios for several model X-ray source
distributions are also plotted in Figure 6.  The upper 3 curves show
models with power law flux distributions with photon index slopes of
1.5, 2.0, and 4.0 respectively, and the 2 lower curves show blackbody
models with temperatures of kT = 0.1 and 0.2 keV.  In each case the
lines connect the calculated SR1 and HR1 values for 4 $N_H$ hydrogen
column densities of, from left to right, $10^{20}$, $5 \times 10^{20}$,
$10^{21}$, and $10^{22}$ atoms cm$^{-2}$.  These models show that the
SR1 color is most sensitive to the $N_H$ parameter such that the more
heavily obscured sources lie towards the right hand side of the
figure.  The HR1 color, on the other hand, is a better indicator of the
intrinsic shape of the spectral flux distribution such that the flatter
spectrum sources lie towards the top of the figure.

The representative models that are plotted in Figure 6, while not
unique, provide an indication of the physical nature of the objects
that lie in different areas of the color -- color diagram.  The very
soft sources in the lower left corner form a distinct subpopulation
which are discussed further in \S\ref{SSS}.  The expected population of
obscured background AGNs seen through the disk of M101 will generally
have power law spectra with photon index slope in the range $ 1.3 <
\Gamma < 4.0$ and total HI columns $> 5 \times 10^{20}$.  There are
more than enough sources within this range of Figure 6 to account for
the expected number of AGNs in the field and the remainder are likely
to be low mass X-ray binaries in M101 itself.  The 6 AGN candidates
located in the interarm regions of M101, previously discussed in
\S\ref{detections}, fall within this region of the figure (although one
of them, source \#39, lies quite close to the 4.0 slope limit) and are
plotted with open triangles.  The sources below the $\Gamma$ = 4.0
powerlaw model line are inconsistent with the spectrum of a typical AGN
and are therefore more likely to be physically associated with M101.

\section{Temporal Variability}   \label{timevar}

Out of the 37 brightest sources which have more than 60 net counts, 4
of them (sources 85, 93, 98, and 104) showed definite intensity
variability during the {\em Chandra} observation ($< 10^{-5}$
probability of being a constant source based on the Kolmogorov-Smirnov
test), and 4 others (sources 51, 66, 71, and 108) are probably variable
(with probabilities ranging from 0.01 -- 0.06 of being a constant
source).  The flux curves for these sources are shown in Figure 7.  Two
of these variable sources (\#93 and \#71) are now brighter than the
{\em ROSAT} detection threshold, but they were not detected in the
previous {\em ROSAT} observations taken in 1992 and 1996. In addition
to these 8 variable sources, there were also 2 variable sources
previously observed with {\em ROSAT} (sources H28 and H31) which
apparently have now faded below the {\em Chandra} detection limit.

Object \#98 is currently the brightest and also the most dramatically
variable source in M101.  It has increased by about a factor of 25 in
mean flux level from when it was previously observed by {\em ROSAT} and the
count rate varied by more than a factor of 6 during the {\em Chandra}
observation alone.

\section{X-ray Models and Fluxes} \label{modelflux}

The 29 brightest sources have enough flux ($> 100$ counts) to perform
at least crude spectral modeling using the HEASARC XSPEC \citep{arn96}
model  fitting program.  Most of the sources could be satisfactorily
fit with a simple absorbed power law spectrum model, however 8 of the
softest sources were better fit by an absorbed blackbody model.   Table
4 lists the best fitting model parameters for these 29 sources.  Also
tabulated are the apparent X-ray fluxes calculated by integrating the
model spectrum over the energy range of the S2 (0.125 -- 0.5 keV), M2
(0.5 -- 2.0 keV), and H2 (2.0 -- 8.0 keV) energy bands as well as the
intrinsic X-ray luminosity of the source in each of the 3 bands,
correcting for the model $N_H$ absorption and assuming a distance of
7.2 Mpc to M101.  In general the models provided a satisfactory fit to
the observed spectra, as shown by the reduced $\chi^2$ values given in
the table.  The main exception to this is the brightest variable source
\#98 whose integrated spectrum could not be well fit by any simple
model.  The X-ray flux from this source is generally very soft, but the
spectrum changes significantly with the flux level.  As a consequence,
the flux and luminosity values given for this object are only
representative of its true nature.  This unusual object will be studied
in more detail in a separate paper, but it is already apparent that
this is a super-Eddington source with an maximum instantaneous
luminosity in the range of $10^{39}$ to $10^{40}$ergs s$^{-1}$

The ratio between the total net X-ray counts and the model X-ray flux
in the S2, M2, and H2 bands for the 29 brightest sources was used to
derive a mean counts-to-flux conversion factor which could be applied
to the fainter sources to estimate their fluxes.  Figure 8 shows the
flux-per-count ratio plotted as a function of the hardness ratio, HR2,
for each source.    The adopted multiplication factors, which convert
from total net counts in our 98.2 ks exposure into units of ergs cm$^{-2}$ s$^{-1}$ are

\[
    C_{S2} = (2.37 \pm 0.39) \times 10^{-17},
\]
\[
    C_{M2} = ((3.5 + 0.9 \times HR2) \pm 0.24) \times 10^{-17} 
\]
and
\[
    C_{H2} = (1.89 \pm 0.34) \times 10^{-16} 
\]
where the M2 band conversion factor has a slight dependence on the
hardness ratio of the source.  These conversion factors were used to
compute the X-ray flux values given in Table 1 for all the sources.
Note that all the X-ray flux and luminosity values (but not the net
counts values) quoted in this paper have been corrected for several
instrumental effects.  Every flux and luminosity value has been
multiplied by a nominal factor of 1.02 (or 1.05 in the case of the
slightly smaller apertures that were used for the 2 nuclear sources) to
correct for the flux that was excluded by the circular integration
apertures.  In addition, 6 of the sources needed a $\sim5$\% correction
factor because they were projected near one of the ``warm'' columns
next to the CCD readout amplifier which were excluded during the
initial data cleaning procedure.  Finally, 2 of the sources were
located near the edge of the chip and required an additional vignetting
correction factor because the deliberate wobbling of the telescope
pointing axis during the exposure caused the source to periodically
move on and off the chip.

Figure 9 shows the distribution of $\log F_x$ for all 110 sources
(using the values from Table 1), and Figure 10 shows the distribution of
$\log L_x$ for the 29 brightest sources from Table 4.  Note that it is
not possible to reliably compute the luminosity for the fainter sources
because the individual $N_H$ values are unknown.   Several of the
sources, especially in the softest band, appear to have luminosities
greater than the Eddington limit of $\sim1.3 \times 10^{38}$ ergs
s$^{-1}$ for a one solar mass object, but the uncertainties in the
calculated luminosities are sometimes quite large.  The brightest S2
band source, \#93, for example, has a high luminosity mainly because
the best fitting model has a very steep spectrum and a large hydrogen
column density of $7.5 \times 10^{21}$ cm$^{-2}$ which leads to a large
correction factor to the apparent flux.  The next most luminous S2 band
source is \#98 which, as mentioned previously, is highly variable and
it's spectrum is not well fit by simple flux models.  It has been
plotted at it's mean {\em Chandra} flux level, but it's instantaneous
luminosity has been observed to vary by at least 2 orders of
magnitude.

In principle, the observed color dependence of the counts-to-flux
conversion factor for the M2 band (Fig.~8) could affect the analysis of
the $\log N-\log S$ relation in \S\ref{lognlogs}.  In that analysis we
assumed a constant conversion factor between counts and flux, but if
the mean X-ray color of the sources changes with count rate (i.e., if
the fainter sources are significantly harder or softer in the mean than
the brighter sources) then this would cause a systematic difference
between the slope of the $\log N-\log(counts)$ and the $\log
N-\log(flux)$ relations.  This difference could be important because
most observational source population studies have used $\log
N-\log(counts)$, whereas most theoretical interpretation is done in
terms of $\log N-\log(flux)$.

In order to judge the importance of this effect, we computed the mean
HR2 color of sources as a function of count rate, as shown in Figure
11.  The fact that there is no significant trend in the figure (the
best fitting line has a slope of $-0.013 \pm 0.16$) means that our
assumption of a constant counts-to-flux conversion factor has not
biased the derived slope of the $\log N-\log S$ relation.

\section{Super Soft Sources}   \label{SSS}

Ten of the {\em Chandra} X-ray sources have very soft spectra similar
to the Supersoft sources (SSS) that have been seen in other nearby
galaxies.  These M101 sources can be divided into 2 distinct
subclasses:  the brighter class consists of 3 sources (13\footnote{
The only source detected in M101 in the {\em Einstein} IPC ultrasoft
survey,  at  $14^h03^m00^s$, $54^{\circ}22$\arcmin25\arcsec 
\citep{tho98} may coincide with our source \#13.  The positions
differ by 1\farcm3, which is about twice the typical error in the {\em
Einstein} source positions, but is still well within the 3\arcmin
~integration radius that was used to extract the {\em Einstein} source
counts.
}, 45, and 99)
and the fainter and softer class consists of 7 sources (8, 16, 30, 72,
80, 96, and 101).  The brighter class of SSS have a mean count rate of
2.0 counts ks$^{-1}$ and have a S2 band (0.125 -- 0.5 keV) to M2 band
count ratio of 3:1.  The fainter class of SSS, on the other hand, are
an order of magnitude fainter with a mean flux of 0.26 counts
ks$^{-1}$, and have such a soft spectrum that essentially all the flux
is emitted in the S2 band.

The 3 brighter SSS have very similar spectra as demonstrated by the good
agreement between the blackbody model fits given in Table 4 for the
sources:  the best fitting model temperatures range from 60 to 100 eV
(equivalent to 0.7 to 1.1 million degrees K) and the hydrogen column
densities lie in the range of 1.0 to $8.0 \times 10^{20}$ cm$^{-2}$,
which is consistent with the Galactic column density of $\sim1.2 \times
10^{20}$  cm$^{-2}$ in the direction of M101 \citep{sta92,har97} plus
an additional absorption component local to the sources in M101.
Because the spectra are so similar,  we combined the counts from the 3
sources to obtain the better signal to noise spectrum shown in Figure
12.  The best fitting blackbody model to this combined spectrum has a
temperature of $72 \pm 2$ eV (equivalent to $ 8.4 \pm 0.2 \times 10^5$
K) and  $N_H = 6.0 \pm 0.7 \times 10^{20}$ atoms cm$^{-2}$.  Note that
the formal errors on the model parameters are unrealistically small,
and do not take into account the unknown systematic errors in the
calibration of the instrumental response below 0.3 kev.  The mean
luminosity of these 3 sources over the 0.125 to 2.0 keV band is $1.4
\times 10^{38}$ ergs s$^{-1}$ with most of this flux emitted below 0.5
keV.  None of these 3 sources showed any obvious signs of variability,
either during the {\em Chandra} exposure or between the epochs of the
{\em ROSAT} and {\em Chandra} observations.

The combined spectrum of the 7 fainter SSS in M101 is shown in Figure
13.  In spite of the statistically large uncertainties in the count
rates and the large systematic uncertainties in the energy calibration
of the {\em Chandra} detector at these low energies, it is clear that
this fainter class of SSS has a much softer spectrum than the brighter
class shown in Figure 12.   Since the  $N_H$ value is not well
constrainted by the model fits, we assumed a conservative value equal
to the Galactic column of $1.2 \times 10^{20}$ cm$^{-2}$.  Under this
assumption the best fitting blackbody model to the mean spectrum has a
temperature of $47 \pm 2$ eV and a 0.125 to 2.0 keV luminosity of $1.1
\times 10^{37}$ erg s$^{-1}$ (85\% of this flux is emitted below 0.3
keV).  Assuming a larger $N_H$ value decreases the best fitting model
temperature only slightly, but it greatly increases the calculated
luminosity (by a factor of 7 for $N_H$ = $6 \times 10^{20}$ cm$^{-2}$)
because even a small increase in column density produces a large
increase in  extinction for these soft sources.

The SSS were first detected in other nearby galaxies by {\em Einstein}
\citep{lon81}, and later studied more extensively with {\em ROSAT}
\citep{tru91, gre91}.  \citet{sup97} classified 3.8\% (15 out of 396)
of the {\em ROSAT} PSPC sources in M31 as being Supersoft sources,
This is not necessarily inconsistent with the higher rate of SSS we
found in M101 (9\%), however, because  the higher Galactic $N_H$
absorption towards M31 ($6 \times 10^{20}$ cm$^{-2}$) compared to M101
will absorb much of the soft flux from these sources.   For this same
reason, the SSS are difficult to detect in our Galaxy if they are more
than $\sim1$ kpc away in the plane of the disk.  Clearly, it will be of
great interest to see how the properties of the SSS population vary
between different types of galaxies as more data from the {\em Chandra}
and XMM observatories become available in the near future.

While the diversity of observed properties argues against lumping all
SSS into a single class of object, the generally accepted model for the
classical SSS is that they are white dwarf stars with steady nuclear
burning in their envelopes \citep{van92}.  It is thought that high
accretion rates ($> 10^{-7} M_{\odot}$ yr$^{-1}$) from a low mass
secondary onto the white dwarf primary fuels the burning of the
hydrogen near the surface of the white dwarf.  The predicted
luminosities range from $6 \times 10^{36}$ ergs s$^{-1}$ to $1 \times 10^{38}$
ergs s$^{-1}$, with peak energy fluxes in the 30 -- 50 eV range, in
good agreement with the sources seen in M101.   The steep drop in flux
above about 0.8 keV seen in Figure 12 is typical of most SSS;  recent
high signal to noise ASCA observations \citep{asa98} have shown that
this break can be explained by an absorption edge feature probably due
to \ion{O}{8} at about 0.87 keV.  This and other edges originating from
hydrogen-like or helium-like ions of carbon, nitrogen and oxygen would
be expected for the emission from a hot white dwarf with a stable
hydrogen burning envelope.

The SSS are also related to classical novae (CNe) in that SSS are
quasi-steady nuclear burning, whereas CNe are the result of
thermonuclear runaways.  Indeed, many CNe go through a SSS phase
several months after the peak of optical outburst.  In one case, the
SSS phase lasted for nearly a decade \citep{sha95}.  We looked for a
similar correlation in M101 by comparing the positions of the 7 known
optical novae that occurred within our M101 field between 1994 and 1997
\citep{sha00} with our X-ray source catalog and with the position of
even fainter possible sources seen in the adaptively smoothed X-ray
image, but found no matches.  Thus, at least none of these novae had a
long-lasting, bright, SSS phase after their optical outburst.

\section{Conclusions}

This paper provides a preview of what the studies of the X-ray source
populations in external galaxies will reveal and what they will teach
us about the X-ray sources in our own Galaxy.  The main result of this
paper is the catalog of 110 X-ray sources seen in the {\em Chandra}
observation of the face-on spiral galaxy M101.  The detection threshold
is about $10^{36}$ ergs s$^{-1}$, but depends on position because of
the variable angular resolution of the X-ray telescope across the field
of view.  Since this flux threshold is several orders of magnitude
higher than the level of coronal X-ray emission expected from normal
stars, most of these sources must be more exotic objects, including
X-ray binaries, supersoft sources (SSS), and supernovae remnants.  The
field of view of the current observation only covers about 50\% of the
prominent spiral arms in M101, so a complete census of all the X-ray
sources visible with {\em Chandra} in M101 at this flux limit could
contain twice as many objects.

Only about 1/4 of the sources are estimated to be background AGNs,  so
most of our sources are physically associated with M101.  The bulk of
the point sources are related to spiral arms, suggesting that the
population of X-ray sources is derived from young objects in regions of
active star formation, and thus may be dominated by High-Mass X-ray
Binaries.  Conversely, spiral arms have higher stellar densities for
all stellar types, so there may not be an over-abundance of High-Mass
X-ray Binaries.  Most of the sources seen in the interarm regions are
probably AGNs because their surface density is close to the expected
number of background sources, and because the X-ray colors of most of
the interarm sources are consistent with that from a power law spectrum
usually seen in AGNs.

At a distance of 7.2 Mpc {\it Chandra} can provide a luminosity
function down to $\sim10^{36}$ ergs s$^{-1}$, so measuring the
luminosity functions for the nearby galaxies will do much to explore
the link between the X-ray source populations and the star-formation
history, and will help to interpret the X-ray source luminosity
functions being developed for the Milky Way.  The derived $\log N-\log
S$ relation for the sources in M101 is nearly linear with a slope of
$-0.80 \pm 0.05$ over the flux range $\sim10^{36}-10^{38}$ ergs
s$^{-1}$.  We  observed no difference in the slope of the relations
between bulge ($R<1\arcmin$) and disk sources
($1\arcmin<R<4.5\arcmin$), however, there are few sources in the bulge
so the uncertainties in the bulge slope are relatively large.  Another
recently published $\log N-\log S$ study similar in scope to the
present work is that of \citet{ten01} for the galaxy M81 (NGC 3031) at
a distance of 3.63 Mpc \citep{fre94}.  M81 is an Sab galaxy with a
substantially larger bulge than M101.  For the 0.2--8.0 keV band
\citet{ten01} find the $\log N-\log S$ of the disk sources to have a
slope of -0.50 over the flux range $4\times10^{36}-2\times10^{39}$ ergs
s$^{-1}$, somewhat flatter than our relation.  Given the difference in
energy range, as well as possible differences in the slope of the AGN
luminosity functions used, the two values are consistent with each
other.  At a limiting flux of $10^{37}$ ergs s$^{-1}$, however, the
surface density of X-ray sources is nearly four times greater for M81
(0.2 kpc$^{-2}$) than for M101 (0.05 kpc$^{-2}$), presumably reflecting
the differences in the star formation rate between the two galaxies.

Eight of the brighter sources show evidence for short term temporal
variation over the 98.2 ks period of our observation, and 2 other
sources previously observed with {\em ROSAT} are no longer visible with
{\em Chandra}.  The brightest {\em Chandra} source is also the most
variable source, varying by more than a factor of 6 during the
observation.  The current mean flux level is about 25 times higher than
when previously observed with {\em ROSAT}.  This source has a soft
spectrum which varies significantly with the flux level.  Simple
spectral models do not provide a very good fit to the spectrum, but it
is clear that it has a super-Eddington luminosity which sometimes
exceeds  $10^{39}$ ergs s$^{-1}$.  This unusual source will be studied
in more detail in a subsequent paper.

The nuclear X-ray source is non-exceptional and looks very much like
the other sources in the field.  It is not spatially resolved, although
there is evidence for enhanced diffuse X-ray emission near the nucleus
(to be studied in a separate paper).   While this source may well be a
weak AGN similar to that detected in most other nearby galaxies
\citep{ho01}, it could also simply be a typical X-ray binary located in
the dense stellar environment close to the nucleus.  There is another
nearly identical X-ray source just 3\farcs1 to the south of the nucleus
which coincides with a loose cluster of bright stars.   A power-law
spectral model gives a resonable fit to both the nuclear source and the
southern companion, resulting in photon indices of 2.02 and 1.56, and
luminosities of $4.2 \times 10^{37}$ ergs s$^{-1}$ and $3.8 \times
10^{37}$ ergs s$^{-1}$ in the 0.5 -- 2.0 keV band, respectively.

About 9\% of the M101 sources have extremely soft spectra similar to
the SSS seen in other nearby galaxies.   The 10 SSS in M101 can be
naturally divided into 2 classes: the 3 brightest SSS have a mean
luminosity of $1.4 \times 10^{38}$ ergs s$^{-1}$ and a model blackbody
temperature of $72 \pm 2$ eV.  The other 7 SSS have an order of
magnitude smaller X-ray count rate, most of which is emitted below 300
eV.  The model temperature and luminosity of these fainter sources is
dependent on the assumed hydrogen column density, but is in the
neighborhood of 50 eV and  $1.1 \times 10^{37}$ erg s$^{-1}$
respectively.  The observed luminosities and temperatures of both these
classes of SSS are within the range predicted by the standard models of
a white dwarf star with steady nuclear burning in its envelope fueled
by accretion from a low mass secondary.


%% Included in this acknowledgments section are examples of the
%% AASTeX hypertext markup commands. Use \url without the optional [HREF]
%% argument when you want to print the url directly in the text. Otherwise,
%% use either \url or \anchor, with the HREF as the first argument and the
%% text to be printed in the second.

\acknowledgments

We wish thank the referee, Dr Philip Kaaret, for his careful review and
useful suggestions that improved this manuscript.  This research was
supported in part by NASA grant number 01900441.  We have made use of
data and software obtained from the High Energy Astrophysics Science
Archive Research Center (HEASARC), provided by NASA's Goddard Space
Flight Center.

\clearpage

\begin{figure}
\plotone{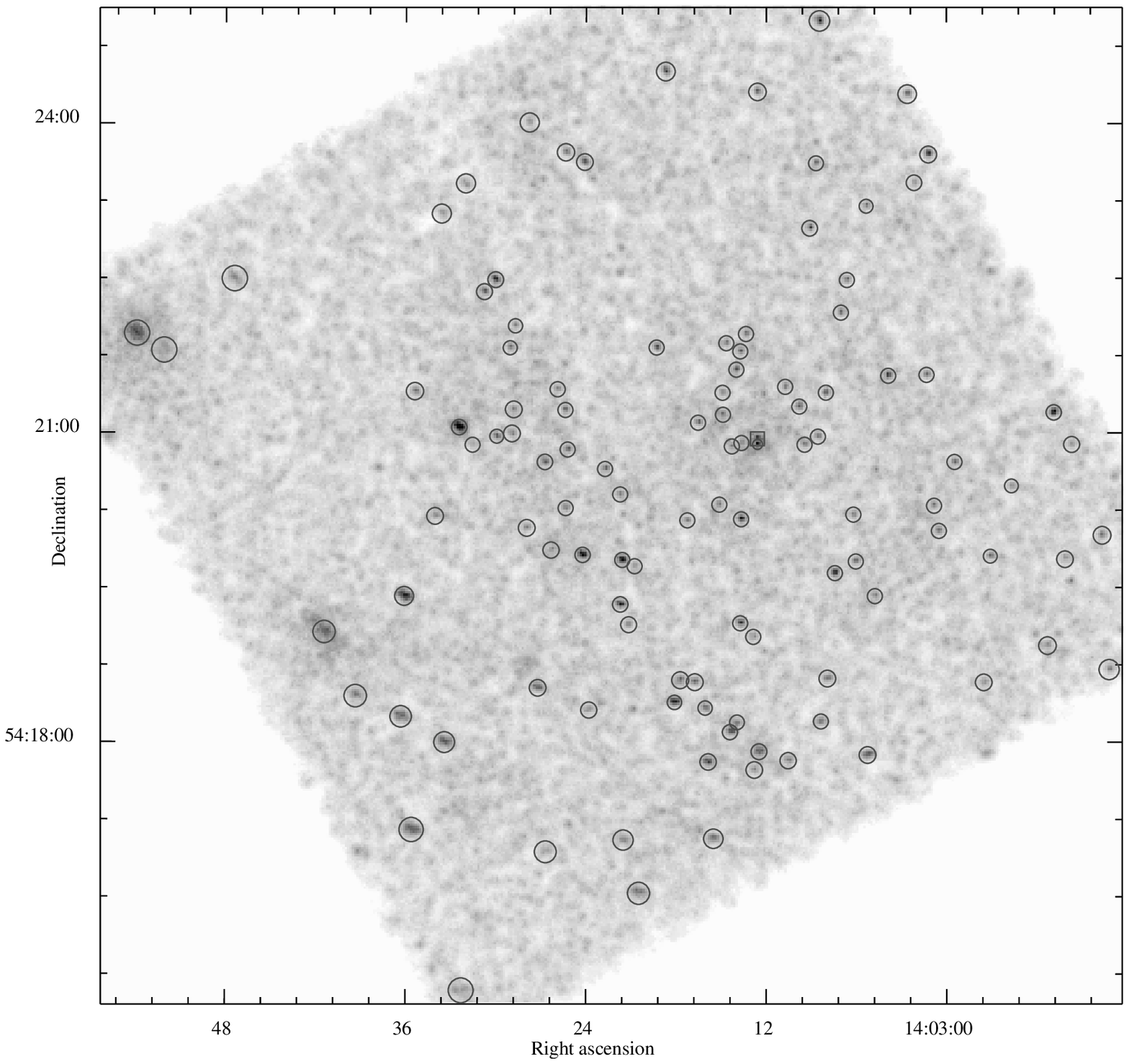}
\caption{Adaptively smoothed {\em Chandra} image of the M101 field covered by
the S3 chip showing the location of the 110 X-ray sources.  The nuclear source
is indicated by the square, to the right and slightly above center.
 \label{fig1}}
\end{figure}

\begin{figure}
\plotone{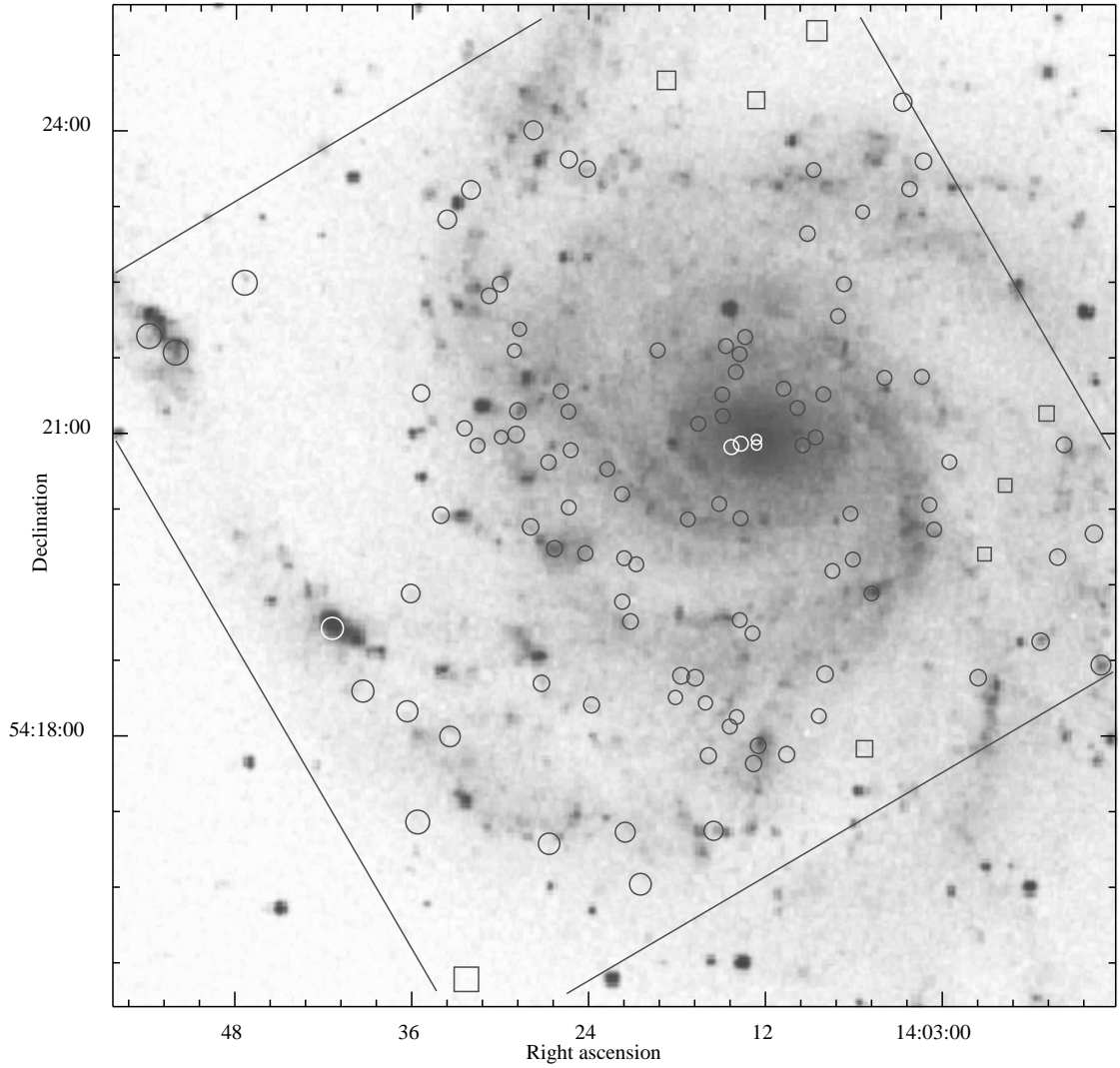}
\caption{Location of the 110 {\em Chandra} X-ray sources plotted on top of
the image of M101 from the digitized Palomar Sky Survey E plate.  It
can be seen that most of the sources are located within the spiral arm
regions.  The squares indicated the 8 ``interarm'' sources that are
discussed in the text.  Some of the regions were shown in white simply
to improve the contrast with the underlying image.
 \label{fig2}}
\end{figure}

\begin{figure}[tb]
\plotone{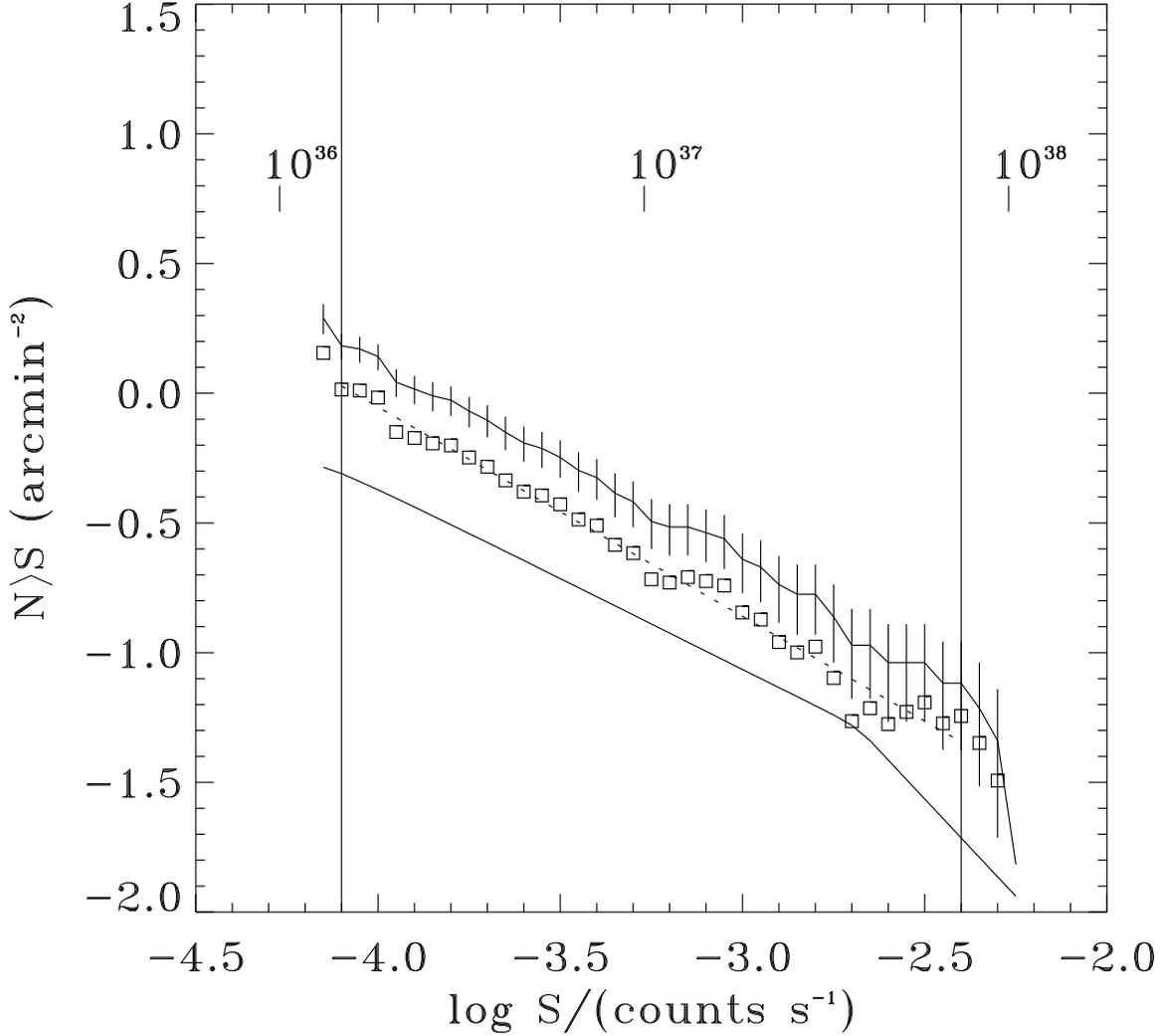}
%%\centerline{\psfig{figure=f3.eps,width=8.0cm}}
\caption{The $\log N-\log S$ relation for M101.
The points with error bars are the relation for all the point sources
with $\sigma>4.5$,
the lower smooth line is the calculated relation
for the background AGN absorbed by the M101 disk,
and the boxes are the difference between those two relations.
The dotted line is the fit; the vertical lines show
the region over which the fit was made.  A flux conversion 
factor equivalent to 
$10^{-14}$ ergs cm$^{-2}$ s$^{-1}$ = $3.33 \times 10^{-3}$ counts s$^{-1}$,
which is appropriate for a background AGN with a $\Gamma=1.42$ power-law X-ray spectrum, was used in these calculations. }

\label{fig:lnls}
\end{figure}

\begin{figure}[b!]
\plotone{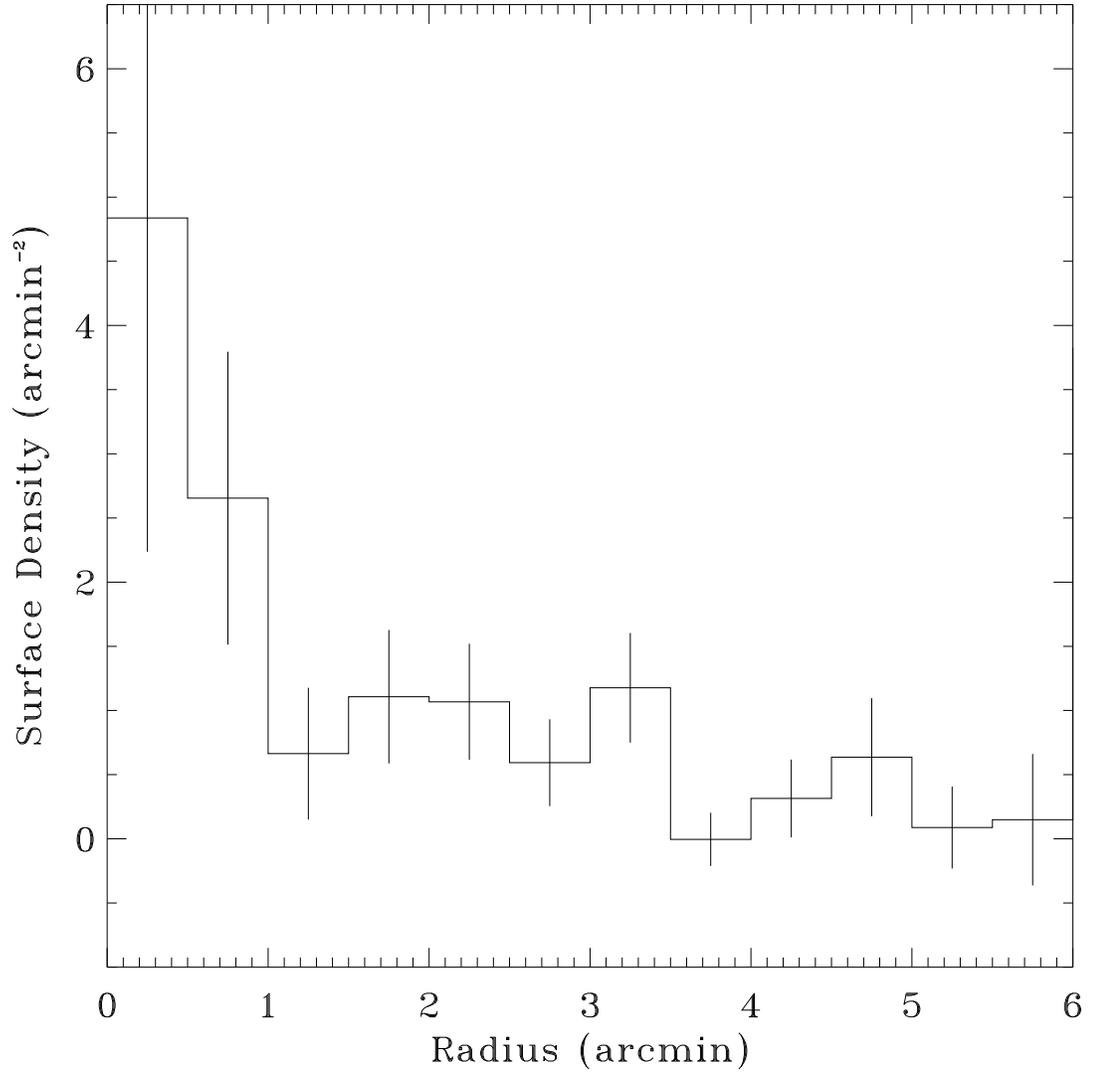}
%%\centerline{\psfig{figure=f4.eps,width=8.0cm}}
\caption{The surface density of point sources as a function
of the distance from the nucleus,
calculated for annuli with $\Delta R=0\farcm5$.}
\label{fig:rad}
\end{figure}

\begin{figure}
\plotone{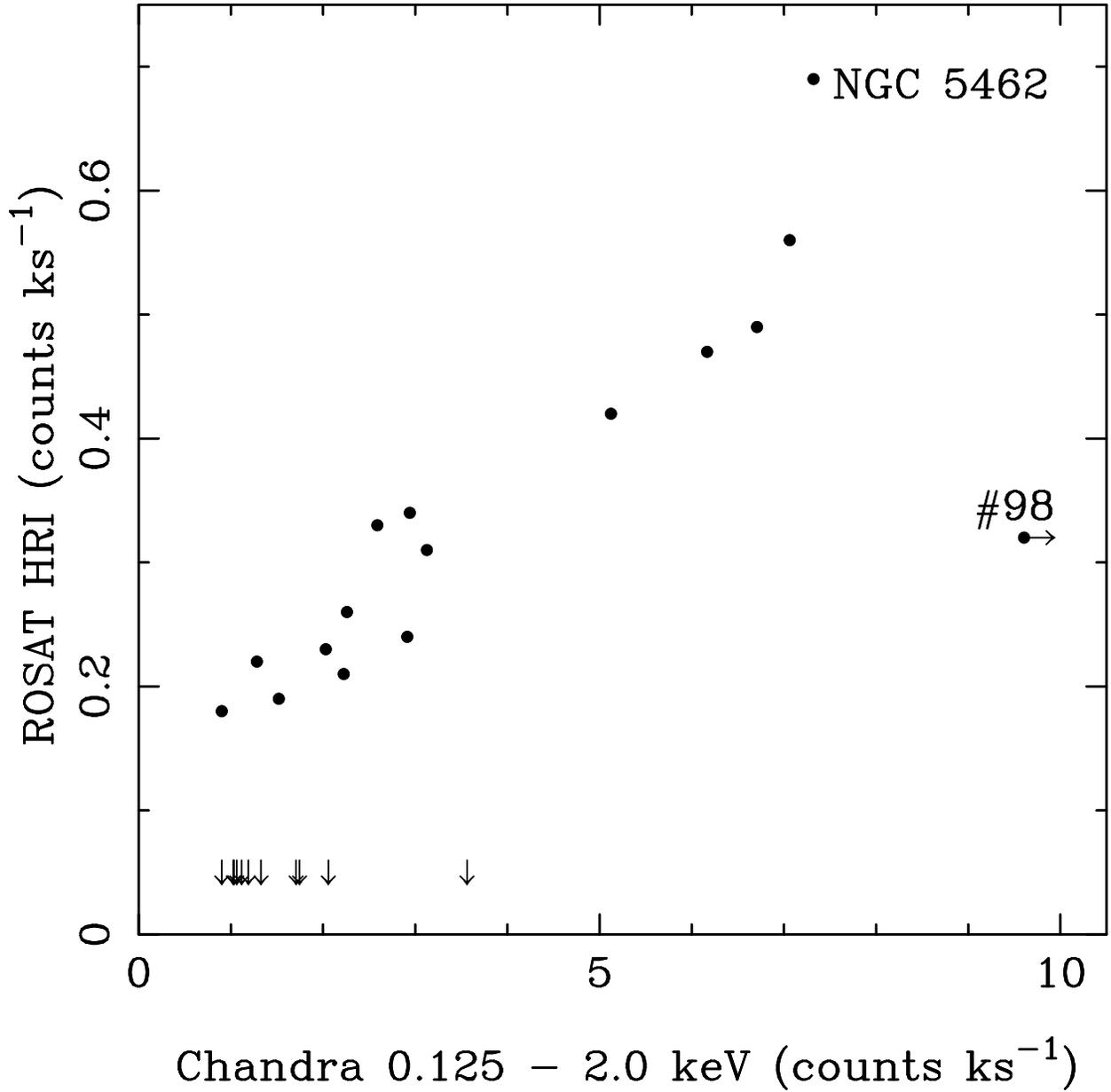}
\caption{Comparison of the {\em Chandra} and {\em ROSAT} count rates.  The bright
source \#98 (plotted at 1/10th of its mean {\em Chandra} count rates in
order to fit it on the graph)  is highly variable and now has a mean flux
level $\sim25$ times brighter than when observed by {\em ROSAT}.  The arrows
at the bottom of the figure show the {\em Chandra} count rates for sources
that were not previously detected by {\em ROSAT}.
 \label{fig5}}
\end{figure}

\begin{figure}
\plotone{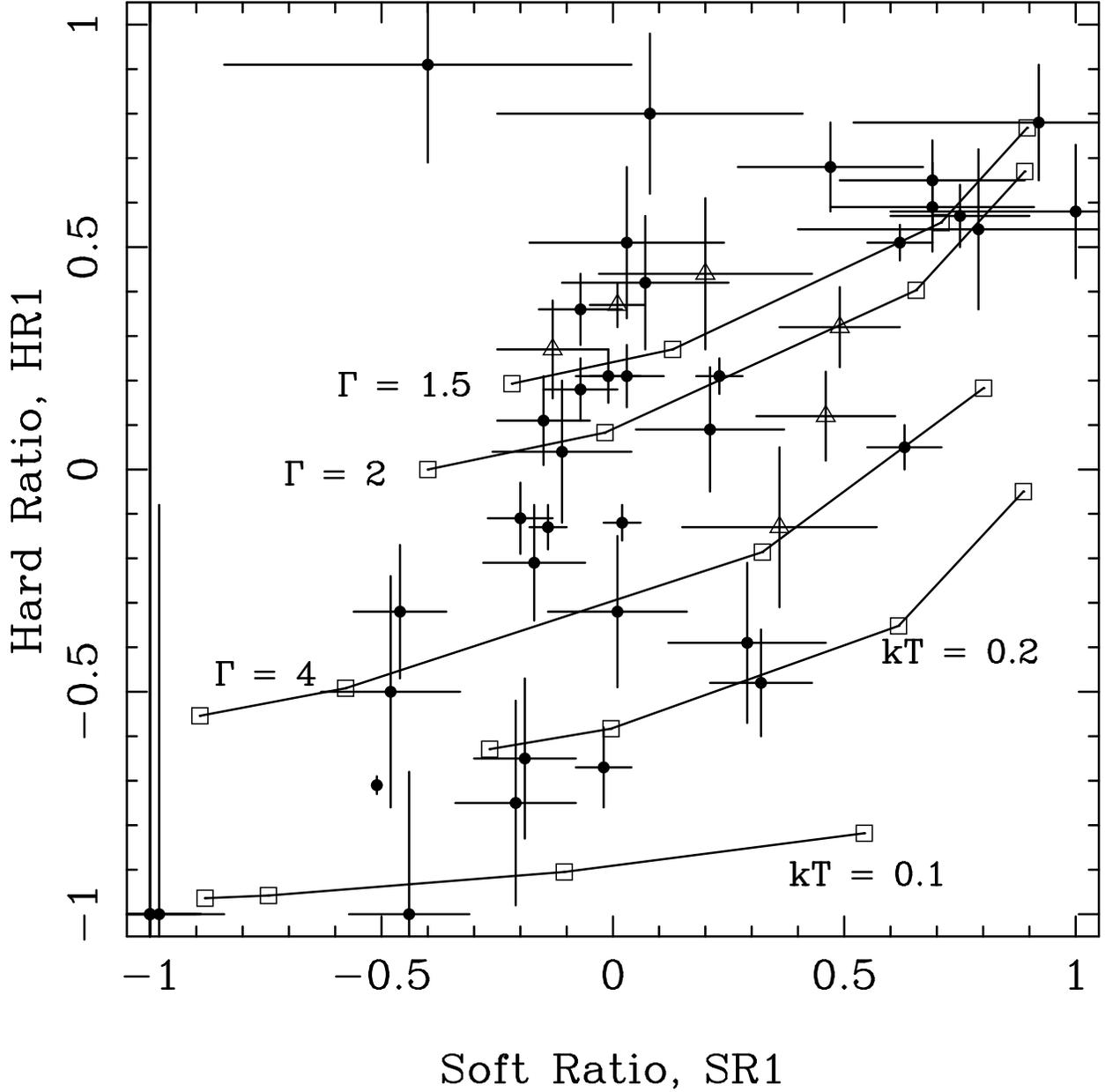}
\caption{Observed X-ray color-color plot for the 45 brightest M101
sources which have more than 40 net X-ray counts,  shown  with the
1-sigma error bars.  The 6 likely AGN sources located in the interarm
regions of M101 are plotted with open triangles.  The lines connecting
the open squares show the predicted colors for 5 different sets of
models.  The upper 3 models have power law flux distributions with the
indicated photon index slopes and the lower 2 models are black bodies
with the indicated temperatures (in units of keV). For each model the
open squares are plotted at $N_H$ values of, from left to right,
$10^{20}$, $5 \times 10^{20}$, $10^{21}$, and $10^{22}$ atoms
cm$^{-2}$.
 \label{fig6}}
\end{figure}

\begin{figure}
\plotone{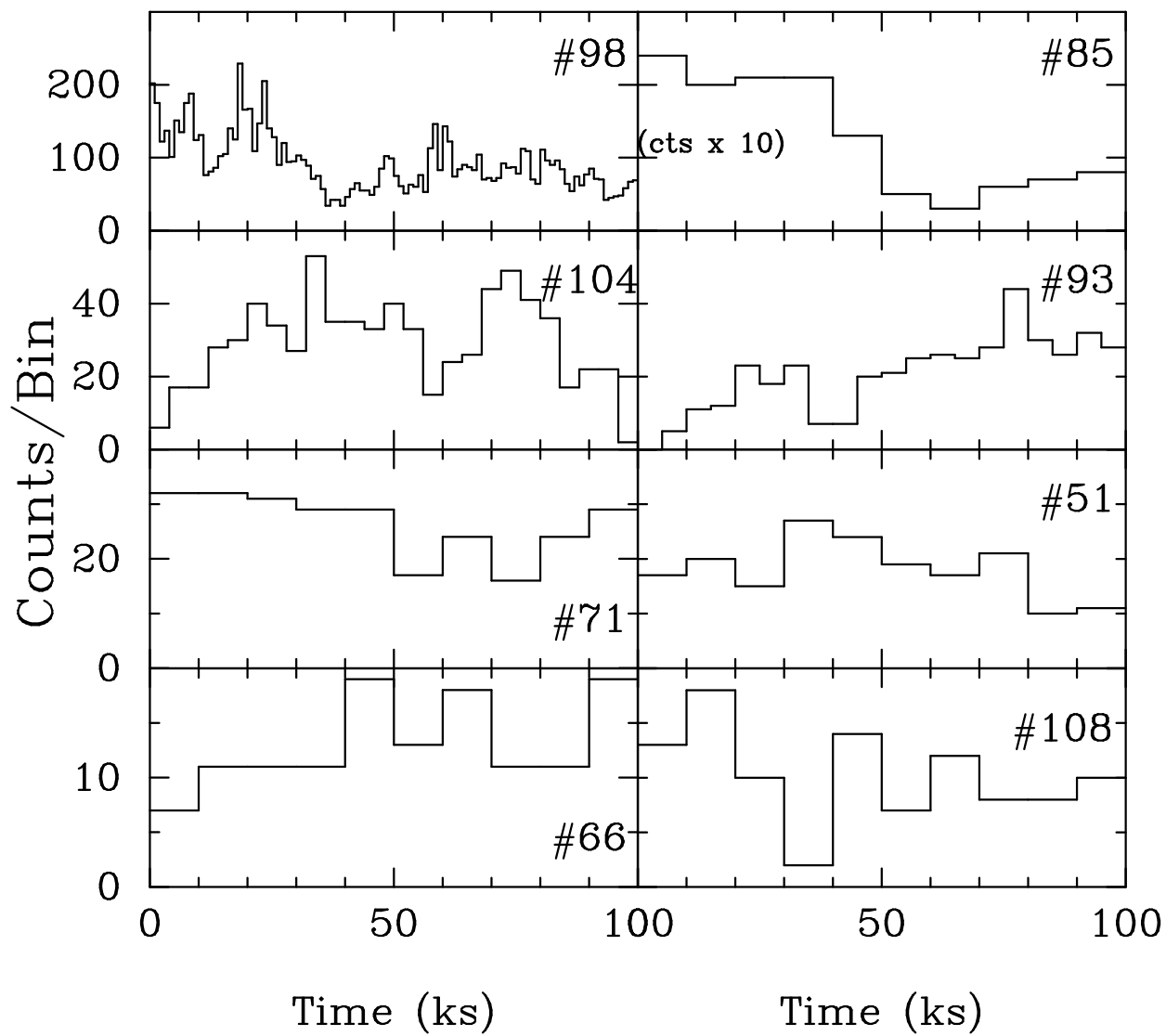}
\caption{X-ray light curves for the variable sources.  The upper 4 sources
are definitely variable, while the lower 4 sources are only marginally
variable according to the Kolmogorov-Smirnov probability test for a
constant source.
 \label{fig7}}
\end{figure}

\begin{figure}
\plotone{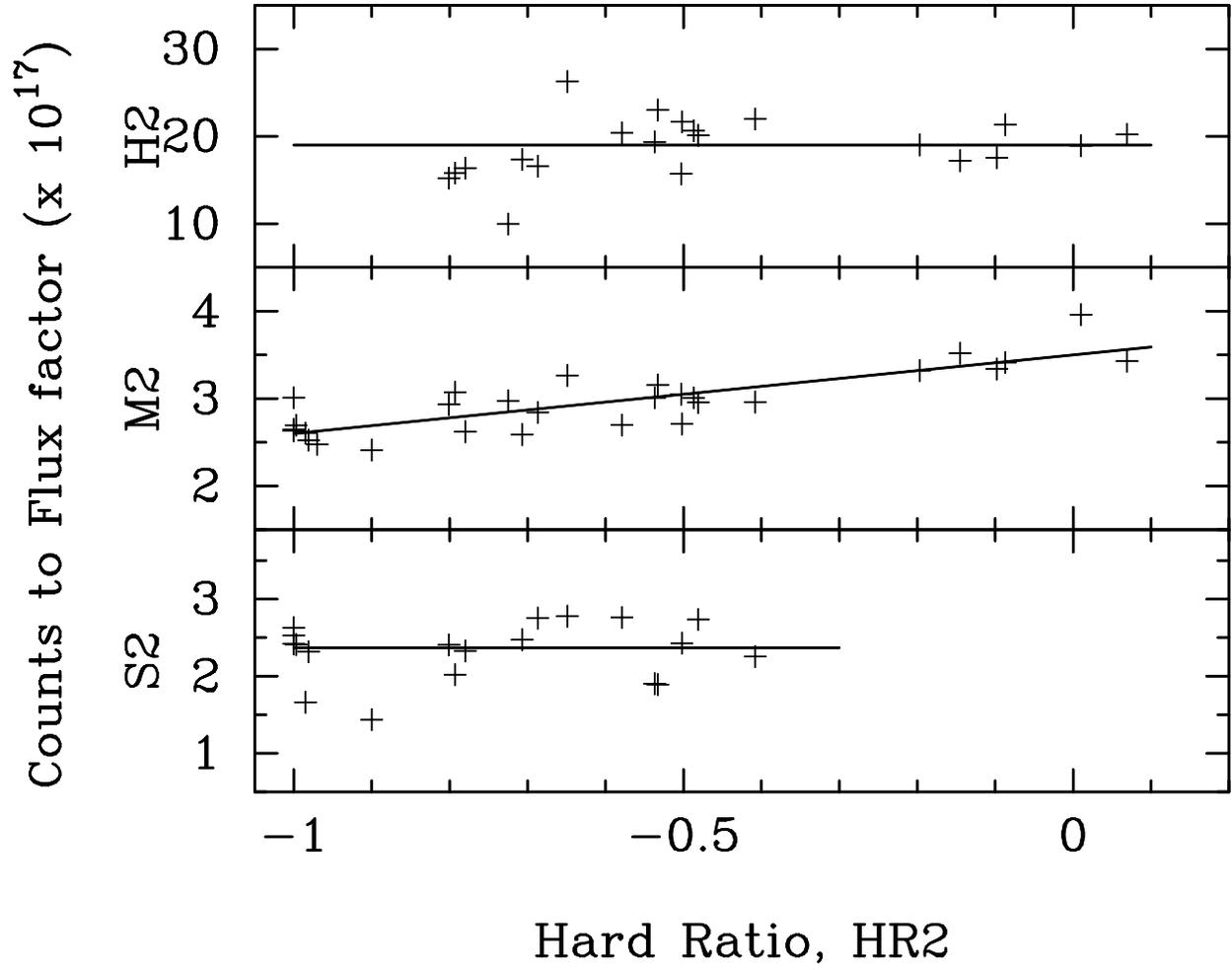}
\caption{Derived counts to flux conversion factors for the S2, M2, and H2
energy bands, based on the model spectra fit to the 29 brightest sources.
 \label{fig8}}
\end{figure}

\begin{figure}
\plotone{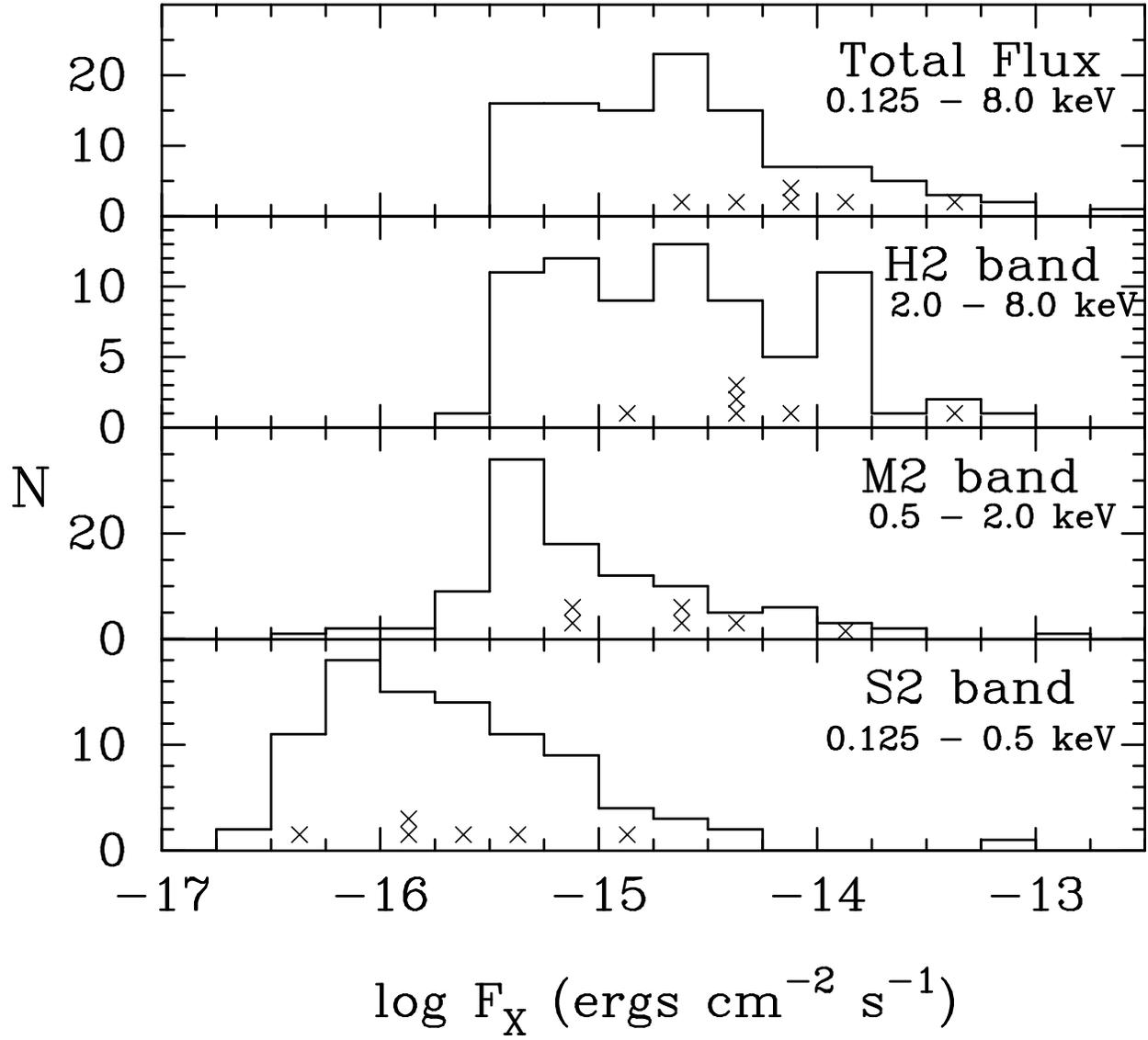}
\caption{X-ray flux histograms for all 110 sources in the 3 energy
bands and the total flux.  The positions in the histograms of the 6
interarm AGN candidates, discussed in \S\ref{detections}, are indicated
by the X symbols.
 \label{fig9}}
\end{figure}

\begin{figure}
\plotone{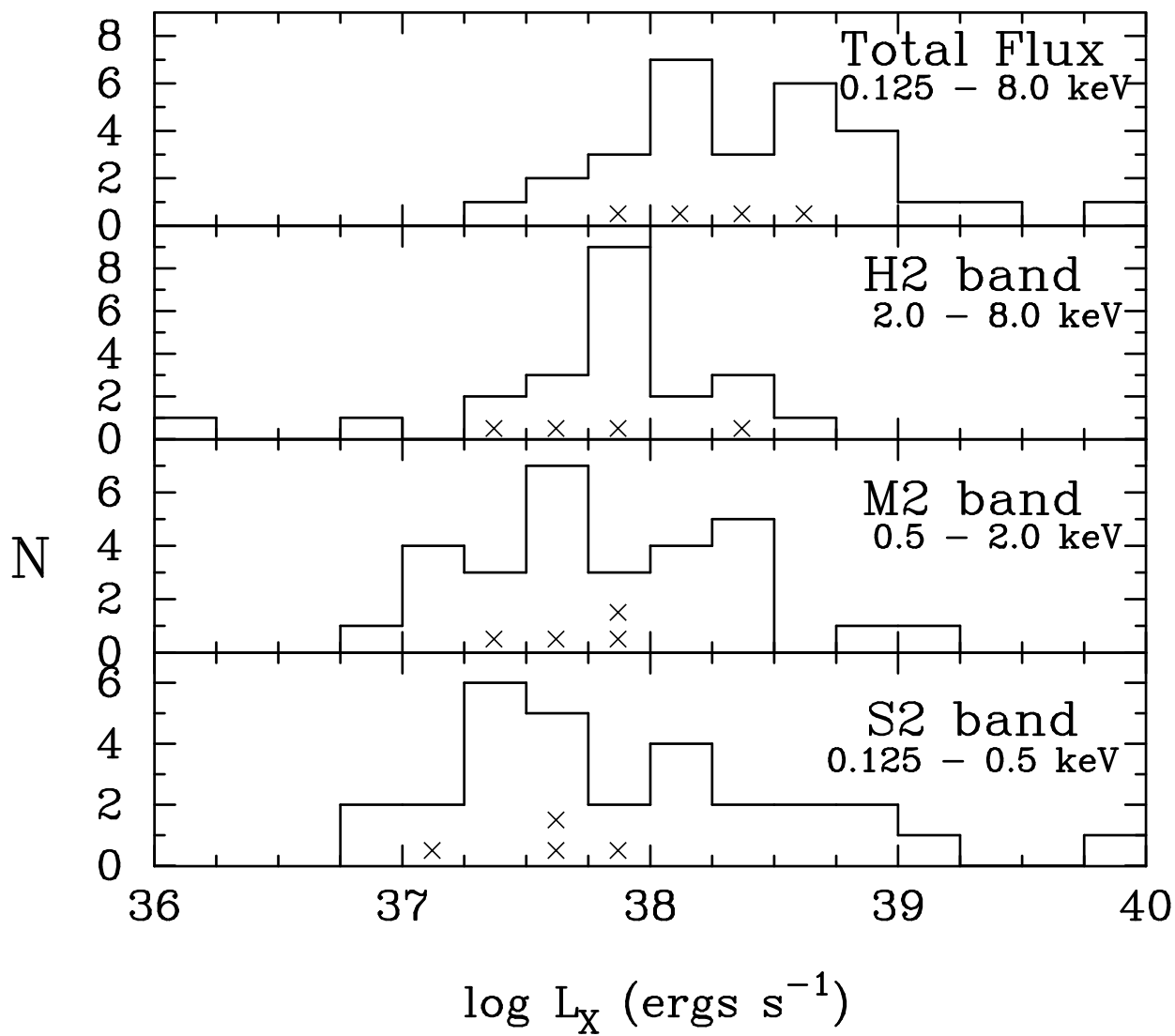}
\caption{ X-ray luminosity histograms for the 29 brightest sources in
the 3 energy bands and the total luminosity.  The positions in the
histograms of the 4 brightest interarm AGN candidates, discussed in
\S\ref{detections}, are indicated by the X symbols.
 \label{fig10}}
\end{figure}

\begin{figure}
\plotone{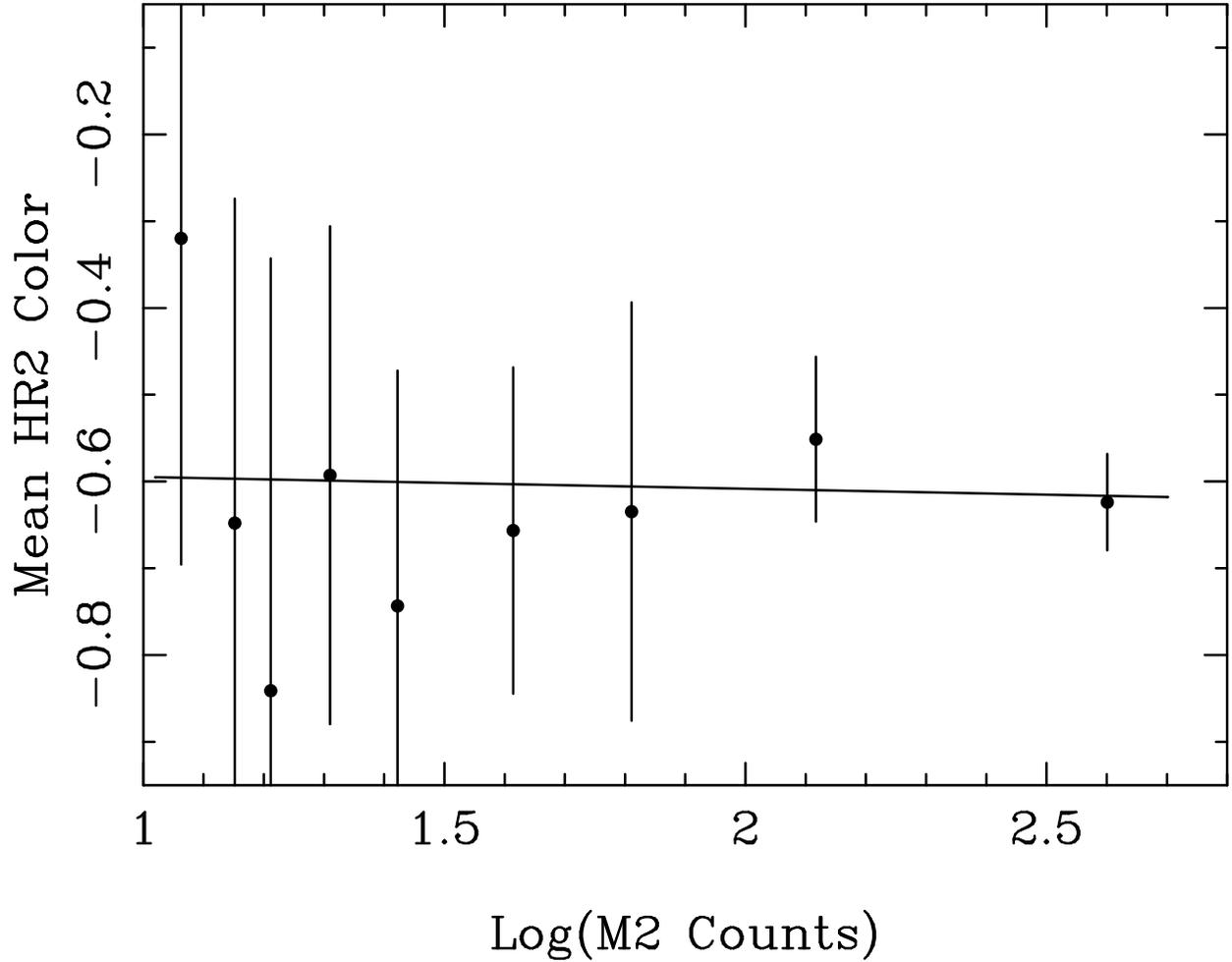}
\caption{ Mean HR2 color as a function of source count rate.  The
sources were sorted in order of M2 band count rate, then the mean HR2
color was computed for bins containing 10 sources each.  The error bars
show the mean statistical error on the HR2 colors for the individual
sources in each bin.  The nearly horizontal line is the least squares
fit to the points, which shows there is no significant trend in mean
HR2 color as a function of count rate.
 \label{fig11}}
\end{figure}

\begin{figure}
\plotone{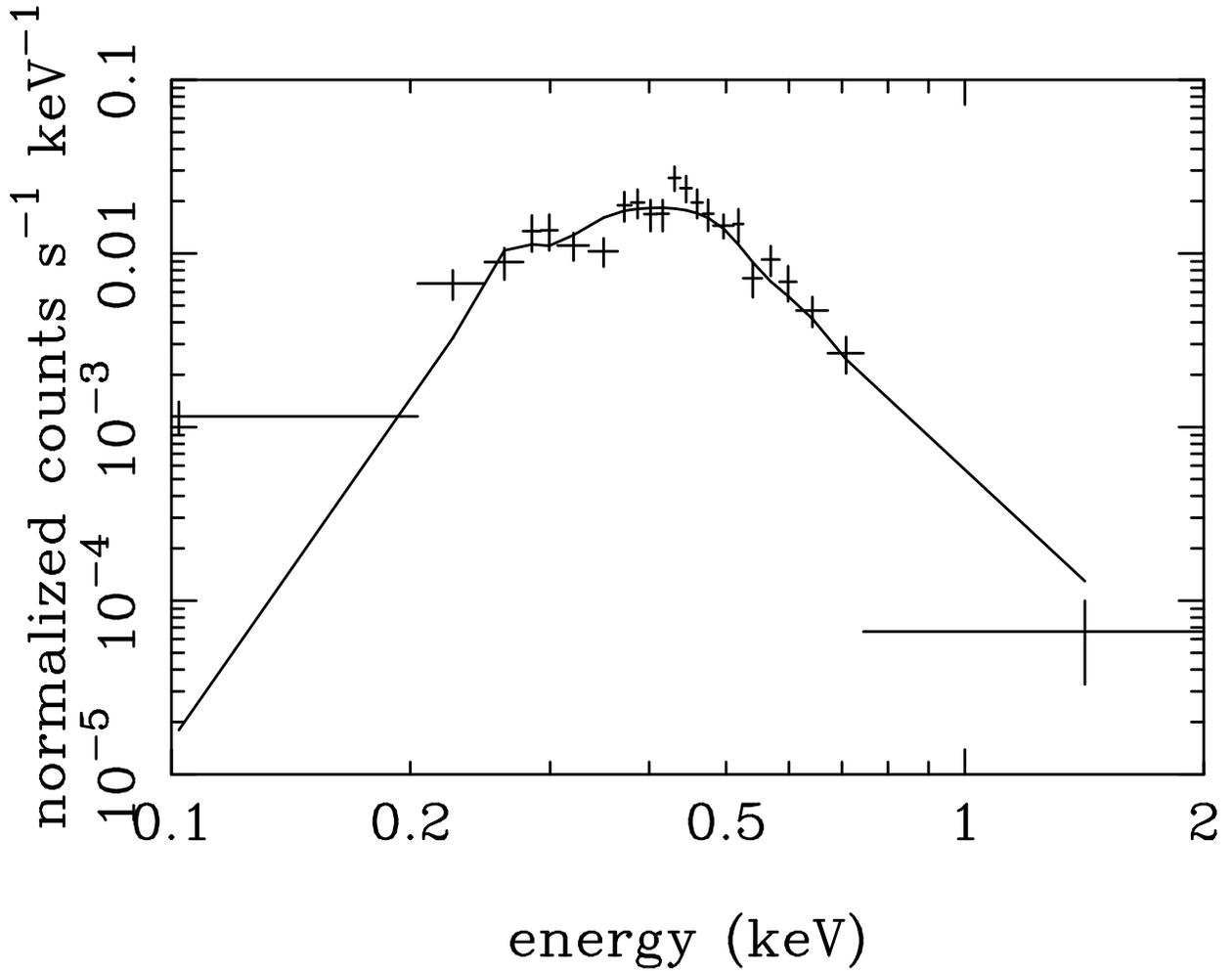}
\caption{Combined X-ray spectrum of the 3 super soft sources cited in the text.
The line shows the best fitting model spectrum with a
blackbody temperature of 72 eV and $N_H$ = $5.6 \times 10^{20}$ cm$^{-2}$.
 \label{fig12}}
\end{figure}

\begin{figure}
\plotone{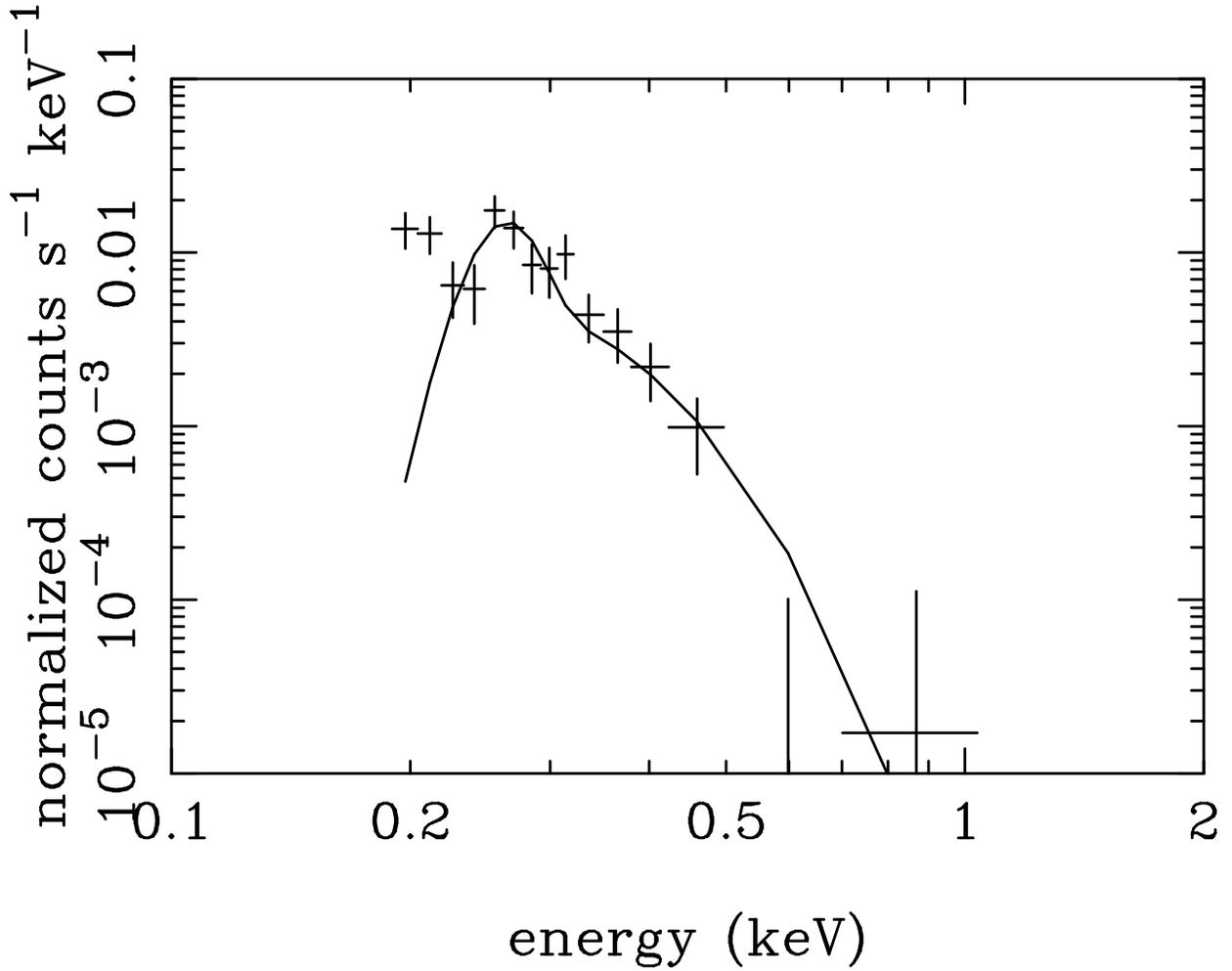}
\caption{Combined X-ray spectrum of the 7 extremely soft sources cited in the text.  The line shows the best fitting model spectrum with a blackbody
temperature of 47 eV and an assumed value of $N_H$ = $1.2 \times
10^{20}$ cm$^{-2}$.
 \label{fig13}}
\end{figure}

%% Tables should be submitted one per page, so put a \clearpage before

%% Two options are available to the author for producing tables:  the
%% deluxetable environment provided by the AASTeX package or the LaTeX
%% table environment.  Use of deluxetable is preferred.
%%

%% Tables may also be prepared as separate files. See the accompanying
%% sample file table.tex for an example of an external table file.
%% To include an external file in your main document, use the \input
%% command. Uncomment the line below to include table.tex in this
%% sample file. (Note that you will need to comment out the \documentclass,
%% \begin{document}, and \end{document} commands from table.tex if you want
%% to include it in this document.)

\input{table1}
\input{table2}
\input{table3}
\input{table4}

%% The following command ends your manuscript. LaTeX will ignore any text

\end{document}

%% file: table1.tex
%\documentclass{aastex}
%\documentclass[preprint]{aastex}
%\begin{document}

\begin{deluxetable}{rcccrrrrrrrrr}
\tablecaption{S3 Chip X-ray Sources in M101}
\tablewidth{0pt}

\tablehead{
\colhead{} & \colhead{}         & \colhead{}     & \colhead{}   & \multicolumn{3}{c}{Net}              & \multicolumn{3}{c}{Net}            & \multicolumn{3}{c}{log($F_X$)} \\
\colhead{} & \colhead{R.A.}     & \colhead{Decl.} & \colhead{Area}  & \multicolumn{3}{c}{Counts}             & \multicolumn{3}{c}{Counts}       & \multicolumn{3}{c}{(ergs cm$^{-2}$ s$^{-1}$)} \\
\colhead{S} & \colhead{(J2000)} & \colhead{(J2000)} & \colhead{(pix)} & \colhead{S1} & \colhead{M1} & \colhead{H1} & \colhead{S2} & \colhead{M2} & \colhead{H2} & \colhead{S2} & \colhead{M2} & \colhead{H2}
}
\startdata

1 & 14 02 49.15 & 54 18 42.1 &   344 &     9 &     5 &     3 &     2 &    18 &    -4 & -16.26 & -15.27 &   \nodata \\
2 & 14 02 49.61 & 54 20 00.2 &   261 &     6 &    12 &     6 &    -4 &    27 &     0 &   \nodata & -15.13 &   \nodata \\
3 & 14 02 51.63 & 54 20 53.1 &   199 &     4 &     3 &     9 &     2 &     8 &     5 & -16.23 & -15.57 & -15.01 \\
4 & 14 02 52.08 & 54 19 46.3 &   224 &    10 &     5 &    -0 &     3 &    11 &     1 & -16.16 & -15.49 & -15.57 \\
5 & 14 02 52.82 & 54 21 11.8 &   185 &   160 &   164 &   355 &    52 &   441 &   185 & -14.90 & -13.85 & -13.45 \\
6 & 14 02 53.25 & 54 18 56.1 &   255 &     4 &    10 &     7 &     4 &    15 &     3 & -16.00 & -15.37 & -15.30 \\
7 & 14 02 55.64 & 54 20 29.0 &   144 &    12 &     1 &     1 &     3 &     9 &     2 & -16.19 & -15.55 & -15.33 \\
8 & 14 02 57.05 & 54 19 48.1 &   142 &    31 &    -1 &    -2 &    32 &    -3 &    -1 & -15.12 &   \nodata &   \nodata \\
9 & 14 02 57.49 & 54 18 34.7 &   223 &    -5 &     6 &    14 &    -4 &     8 &    11 &   \nodata & -15.49 & -14.64 \\
10 & 14 02 59.43 & 54 20 43.0 &    90 &    19 &    12 &     1 &     7 &    25 &     1 & -15.78 & -15.17 &   \nodata \\
11 & 14 03 00.47 & 54 20 02.9 &    94 &    24 &    10 &     2 &     9 &    25 &     3 & -15.67 & -15.16 & -15.29 \\
12 & 14 03 00.79 & 54 20 17.5 &    88 &    19 &     1 &    -1 &    12 &     8 &    -0 & -15.54 & -15.68 &   \nodata \\
13 & 14 03 01.17 & 54 23 42.0 &   229 &   169 &     0 &    -4 &   124 &    44 &    -3 & -14.52 & -14.93 &   \nodata \\
14 & 14 03 01.29 & 54 21 33.7 &    90 &     8 &    10 &     1 &     2 &    18 &    -0 & -16.33 & -15.33 &   \nodata \\
15 & 14 03 02.12 & 54 23 25.5 &   192 &    -4 &     9 &     5 &    -1 &     8 &     3 &   \nodata & -15.59 & -15.21 \\
16 & 14 03 02.57 & 54 24 17.1 &   280 &    43 &    -1 &    -3 &    46 &    -2 &    -4 & -14.82 &   \nodata &   \nodata \\
17 & 14 03 03.84 & 54 21 33.2 &    95 &     5 &    26 &   102 &     3 &    61 &    70 & -16.17 & -14.66 & -13.87 \\
18 & 14 03 04.73 & 54 19 24.9 &    88 &     7 &     9 &     2 &     4 &    14 &     1 & -16.02 & -15.42 &   \nodata \\
19 & 14 03 05.22 & 54 17 52.4 &   221 &    40 &    31 &    55 &    16 &    91 &    19 & -15.41 & -14.57 & -14.43 \\
20 & 14 03 05.31 & 54 23 11.9 &   143 &     7 &     8 &     0 &    -1 &    16 &     0 &   \nodata & -15.35 &   \nodata \\
21 & 14 03 06.00 & 54 19 45.1 &    92 &     4 &     0 &    22 &     2 &    10 &    15 & -16.34 & -15.44 & -14.55 \\
22 & 14 03 06.16 & 54 20 12.3 &    94 &     1 &     1 &    12 &     2 &     2 &    11 & -16.35 & -16.15 & -14.69 \\
23 & 14 03 06.59 & 54 22 28.9 &    91 &     4 &     7 &    15 &     4 &    14 &     9 & -16.02 & -15.34 & -14.77 \\
24 & 14 03 06.99 & 54 22 09.9 &    95 &     1 &     1 &    25 &    -1 &    13 &    16 &   \nodata & -15.33 & -14.52 \\
25 & 14 03 07.39 & 54 19 38.3 &    95 &     9 &    62 &   230 &     2 &   163 &   137 & -16.36 & -14.25 & -13.58 \\
26 & 14 03 07.89 & 54 18 36.9 &   137 &    12 &     9 &    17 &     7 &    24 &     8 & -15.78 & -15.14 & -14.84 \\
27 & 14 03 07.99 & 54 21 23.3 &    93 &    28 &     9 &    -1 &    15 &    23 &    -1 & -15.42 & -15.19 &   \nodata \\
28 & 14 03 08.32 & 54 18 12.0 &   170 &     4 &     2 &    38 &     3 &     6 &    35 & -16.13 & -15.61 & -14.17 \\
29 & 14 03 08.42 & 54 24 59.7 &   359 &    18 &    51 &   100 &     9 &   119 &    41 & -15.67 & -14.43 & -14.10 \\
30 & 14 03 08.52 & 54 20 57.7 &    95 &    29 &     0 &    -0 &    27 &     2 &     1 & -15.19 & -16.29 &   \nodata \\
31 & 14 03 08.65 & 54 23 36.8 &   168 &    18 &     3 &     1 &    12 &    10 &    -0 & -15.53 & -15.59 &   \nodata \\
32 & 14 03 09.07 & 54 22 59.0 &   104 &     3 &     6 &    20 &     1 &    21 &     7 &   \nodata & -15.17 & -14.85 \\
33 & 14 03 09.41 & 54 20 53.0 &    92 &     4 &    11 &     8 &     4 &    15 &     5 & -16.03 & -15.34 & -15.04 \\
34 & 14 03 09.76 & 54 21 15.2 &    95 &    11 &    11 &    35 &     5 &    38 &    15 & -15.93 & -14.92 & -14.55 \\
35 & 14 03 10.50 & 54 17 49.1 &   209 &     6 &     6 &    16 &     3 &    15 &    10 & -16.09 & -15.28 & -14.70 \\
36 & 14 03 10.70 & 54 21 26.7 &    94 &    15 &     3 &    -0 &     5 &    14 &    -0 & -15.93 & -15.44 &   \nodata \\
37 & 14 03 12.44 & 54 17 54.3 &   198 &    82 &    31 &    16 &    27 &    96 &     5 & -15.18 & -14.58 & -15.01 \\
38 & 14 03 12.54 & 54 20 53.2 &    60 &    95 &    83 &   119 &    38 &   204 &    54 & -15.03 & -14.19 & -13.97 \\
39 & 14 03 12.55 & 54 24 18.4 &   250 &     8 &    18 &    14 &     4 &    31 &     6 & -15.99 & -15.04 & -14.96 \\
40 & 14 03 12.55 & 54 20 56.5 &    60 &   143 &    95 &    76 &    68 &   208 &    38 & -14.77 & -14.20 & -14.12 \\
41 & 14 03 12.75 & 54 17 43.7 &   222 &     7 &     3 &     6 &     2 &    15 &    -1 & -16.34 & -15.39 &   \nodata \\
42 & 14 03 12.82 & 54 19 01.2 &    88 &    12 &     3 &     0 &     6 &     9 &     1 & -15.84 & -15.61 &   \nodata \\
43 & 14 03 13.31 & 54 21 57.5 &    93 &    26 &     4 &     1 &    13 &    17 &     2 & -15.51 & -15.33 & -15.49 \\
44 & 14 03 13.61 & 54 20 54.0 &    88 &     7 &     3 &     8 &     2 &    12 &     5 & -16.32 & -15.43 & -15.03 \\
45 & 14 03 13.63 & 54 20 09.6 &    94 &   216 &     2 &    -2 &   170 &    48 &    -1 & -14.39 & -14.90 &   \nodata \\
46 & 14 03 13.69 & 54 21 47.3 &    95 &     5 &     1 &    30 &     2 &    12 &    23 & -16.36 & -15.35 & -14.36 \\
47 & 14 03 13.69 & 54 19 09.1 &    93 &    32 &    62 &    22 &     3 &   112 &     2 & -16.16 & -14.52 & -15.49 \\
48 & 14 03 13.90 & 54 18 11.4 &   165 &    23 &    14 &     1 &    11 &    26 &     1 & -15.57 & -15.15 &   \nodata \\
49 & 14 03 13.95 & 54 21 36.7 &    88 &    24 &    19 &    21 &    12 &    46 &     7 & -15.54 & -14.88 & -14.88 \\
50 & 14 03 14.25 & 54 20 52.0 &    93 &     8 &     5 &     2 &     1 &    12 &     3 &   \nodata & -15.46 & -15.29 \\
51 & 14 03 14.37 & 54 18 05.7 &   180 &    63 &    47 &    58 &    21 &   126 &    22 & -15.30 & -14.44 & -14.38 \\
52 & 14 03 14.62 & 54 21 52.1 &    92 &    14 &     7 &    -1 &     8 &    13 &    -0 & -15.72 & -15.47 &   \nodata \\
53 & 14 03 14.84 & 54 21 10.5 &    92 &    13 &     7 &     6 &     3 &    20 &     4 & -16.16 & -15.24 & -15.15 \\
54 & 14 03 14.87 & 54 21 23.2 &    94 &     2 &     7 &     5 &    -0 &    14 &     1 &   \nodata & -15.43 &   \nodata \\
55 & 14 03 15.08 & 54 20 18.1 &    94 &    19 &     1 &     1 &     7 &    14 &     1 & -15.78 & -15.43 &   \nodata \\
56 & 14 03 15.47 & 54 17 03.7 &   320 &    39 &    26 &     4 &     9 &    60 &     0 & -15.68 & -14.80 &   \nodata \\
57 & 14 03 15.82 & 54 17 48.4 &   219 &     6 &    30 &   146 &    -1 &   101 &    83 &   \nodata & -14.46 & -13.80 \\
58 & 14 03 16.02 & 54 18 19.8 &   158 &    14 &    25 &    11 &     0 &    46 &     3 &   \nodata & -14.89 & -15.23 \\
59 & 14 03 16.50 & 54 21 05.8 &    94 &    12 &     2 &    14 &     6 &    13 &    10 & -15.85 & -15.36 & -14.73 \\
60 & 14 03 16.71 & 54 18 34.8 &   133 &    17 &     9 &     7 &     8 &    23 &     3 & -15.72 & -15.19 & -15.29 \\
61 & 14 03 17.21 & 54 20 09.0 &    94 &    15 &     1 &     4 &     7 &     9 &     5 & -15.78 & -15.55 & -15.05 \\
62 & 14 03 17.68 & 54 18 35.9 &   138 &     4 &     5 &    47 &     1 &    21 &    35 &   \nodata & -15.11 & -14.18 \\
63 & 14 03 18.06 & 54 18 23.0 &   157 &   105 &   103 &   157 &    40 &   240 &    84 & -15.01 & -14.12 & -13.79 \\
64 & 14 03 18.66 & 54 24 30.2 &   293 &    15 &    41 &    53 &     1 &    81 &    27 & -16.49 & -14.58 & -14.26 \\
65 & 14 03 19.26 & 54 21 49.6 &    93 &    16 &    24 &    29 &     7 &    43 &    20 & -15.78 & -14.86 & -14.42 \\
66 & 14 03 20.44 & 54 16 31.9 &   427 &     0 &    11 &    88 &     2 &    48 &    49 & -16.26 & -14.74 & -14.00 \\
67 & 14 03 20.72 & 54 19 42.3 &    94 &     6 &    10 &     2 &     1 &    16 &     2 &   \nodata & -15.35 & -15.50 \\
68 & 14 03 21.11 & 54 19 08.2 &   110 &    16 &     3 &    -1 &     2 &    17 &    -1 & -16.22 & -15.34 &   \nodata \\
69 & 14 03 21.47 & 54 17 02.9 &   349 &     1 &    10 &    32 &    -1 &    22 &    21 &   \nodata & -15.10 & -14.39 \\
70 & 14 03 21.54 & 54 19 45.9 &    88 &   174 &   275 &   423 &    40 &   640 &   193 & -15.01 & -13.71 & -13.43 \\
71 & 14 03 21.67 & 54 19 20.0 &   102 &    70 &    73 &   113 &    25 &   173 &    57 & -15.22 & -14.27 & -13.96 \\
72 & 14 03 21.68 & 54 20 24.0 &    94 &    21 &    -1 &    -2 &    20 &     1 &    -2 & -15.32 &   \nodata &   \nodata \\
73 & 14 03 22.69 & 54 20 38.9 &    93 &    29 &     8 &     2 &    11 &    28 &     1 & -15.58 & -15.13 &   \nodata \\
74 & 14 03 23.76 & 54 18 18.5 &   207 &    -0 &    -1 &    20 &    -1 &     9 &    11 &   \nodata & -15.49 & -14.68 \\
75 & 14 03 24.05 & 54 23 37.5 &   227 &    50 &    19 &    -1 &    16 &    54 &    -2 & -15.42 & -14.84 &   \nodata \\
76 & 14 03 24.18 & 54 19 48.9 &    94 &    53 &   228 &   704 &    10 &   584 &   392 & -15.62 & -13.70 & -13.12 \\
77 & 14 03 25.18 & 54 20 50.2 &    94 &     8 &    11 &    21 &     3 &    27 &    11 & -16.16 & -15.07 & -14.69 \\
78 & 14 03 25.31 & 54 23 43.2 &   246 &    -1 &     5 &    21 &    -4 &    14 &    15 &   \nodata & -15.30 & -14.54 \\
79 & 14 03 25.31 & 54 20 16.1 &    91 &     3 &     8 &    13 &     3 &    15 &     7 & -16.15 & -15.32 & -14.89 \\
80 & 14 03 25.33 & 54 21 13.2 &    92 &    29 &     0 &     3 &    28 &     1 &     4 & -15.17 &   \nodata & -15.15 \\
81 & 14 03 25.85 & 54 21 25.2 &    98 &    15 &     4 &    -1 &     3 &    17 &    -1 & -16.18 & -15.36 &   \nodata \\
82 & 14 03 26.28 & 54 19 51.6 &   119 &    14 &     5 &    -1 &     1 &    17 &     0 & -16.52 & -15.34 &   \nodata \\
83 & 14 03 26.64 & 54 16 56.0 &   414 &    18 &     4 &    -1 &     8 &    16 &    -3 & -15.69 & -15.38 &   \nodata \\
84 & 14 03 26.69 & 54 20 42.9 &   102 &    41 &    14 &     5 &    14 &    41 &     4 & -15.48 & -14.93 & -15.07 \\
85 & 14 03 27.16 & 54 18 31.4 &   223 &    51 &    36 &    24 &    26 &    76 &     9 & -15.20 & -14.66 & -14.74 \\
86 & 14 03 27.73 & 54 24 00.5 &   308 &     1 &     4 &    11 &    -1 &     5 &    11 &   \nodata & -15.68 & -14.66 \\
87 & 14 03 27.90 & 54 20 04.5 &   130 &     2 &     2 &    14 &    -1 &    11 &     8 &   \nodata & -15.43 & -14.83 \\
88 & 14 03 28.66 & 54 22 02.1 &   155 &     8 &     6 &     2 &     1 &    12 &     2 & -16.46 & -15.45 & -15.39 \\
89 & 14 03 28.77 & 54 21 13.5 &   130 &    15 &     6 &     5 &     5 &    17 &     4 & -15.92 & -15.30 & -15.14 \\
90 & 14 03 28.89 & 54 20 59.5 &   137 &    12 &    14 &     2 &     5 &    25 &    -1 & -15.93 & -15.19 &   \nodata \\
91 & 14 03 29.01 & 54 21 49.4 &   152 &     8 &    11 &     2 &    -1 &    19 &     3 &   \nodata & -15.25 & -15.21 \\
92 & 14 03 29.90 & 54 20 57.9 &   143 &    78 &    -0 &     0 &    46 &    32 &     0 & -14.96 & -15.06 &   \nodata \\
93 & 14 03 29.98 & 54 22 29.0 &   193 &    38 &   170 &   189 &     3 &   340 &    54 & -16.21 & -14.01 & -13.98 \\
94 & 14 03 30.73 & 54 22 22.0 &   205 &     0 &    13 &    47 &     1 &    35 &    24 & -16.51 & -14.93 & -14.34 \\
95 & 14 03 31.51 & 54 20 52.9 &   169 &     6 &     4 &     3 &     5 &     5 &     3 & -15.91 & -15.80 & -15.27 \\
96 & 14 03 31.97 & 54 23 25.0 &   303 &    21 &     2 &    -7 &    19 &     2 &    -6 & -15.29 & -16.14 &   \nodata \\
97 & 14 03 32.25 & 54 15 35.3 &   736 &     8 &    12 &    30 &     6 &    27 &    18 & -15.83 & -15.05 & -14.47 \\
98 & 14 03 32.40 & 54 21 03.1 &   187 &  6723 &  2210 &   375 &  2828 &  6417 &    62 & -13.17 & -12.77 & -13.92 \\
99 & 14 03 33.37 & 54 17 59.8 &   371 &   215 &    -1 &    -5 &   145 &    69 &    -4 & -14.45 & -14.74 &   \nodata \\
100 & 14 03 33.58 & 54 23 07.4 &   301 &     3 &    -0 &    12 &    -2 &     3 &    13 &   \nodata & -15.85 & -14.59 \\
101 & 14 03 34.00 & 54 20 11.4 &   227 &    20 &    -1 &    -1 &    22 &    -4 &     0 & -15.28 &   \nodata &   \nodata \\
102 & 14 03 35.35 & 54 21 24.1 &   244 &     4 &     5 &    20 &     2 &    20 &     7 & -16.24 & -15.21 & -14.88 \\
103 & 14 03 35.56 & 54 17 08.9 &   534 &    67 &    58 &   123 &    23 &   173 &    53 & -15.26 & -14.27 & -13.99 \\
104 & 14 03 36.05 & 54 19 24.8 &   300 &   303 &   227 &   174 &   117 &   528 &    58 & -14.55 & -13.82 & -13.95 \\
105 & 14 03 36.27 & 54 18 14.8 &   401 &     8 &    23 &   120 &    -0 &    87 &    65 &   \nodata & -14.52 & -13.90 \\
106 & 14 03 39.29 & 54 18 26.7 &   446 &    14 &    16 &    39 &     7 &    44 &    18 & -15.79 & -14.86 & -14.46 \\
107 & 14 03 41.37 & 54 19 04.0 &   437 &   140 &   136 &    27 &    23 &   278 &     2 & -15.26 & -14.13 & -15.40 \\
108 & 14 03 47.36 & 54 22 29.6 &   581 &    23 &    24 &    12 &     8 &    45 &     6 & -15.73 & -14.88 & -14.90 \\
109 & 14 03 52.05 & 54 21 48.0 &   755 &    54 &    37 &     8 &    23 &    76 &     0 & -15.26 & -14.69 &   \nodata \\
110 & 14 03 53.86 & 54 21 57.8 &   756 &   271 &   283 &   223 &    80 &   625 &    72 & -14.72 & -13.75 & -13.86 \\

\enddata
\end{deluxetable}

\clearpage

%\end{document}

%% file: table2.tex
%\documentclass{aastex}
%\documentclass[preprint]{aastex}
%\begin{document}

\begin{deluxetable}{rlll}
\tablecaption{Cross Identifications}
\tablewidth{0pt}

\tablehead{
\colhead{S}  & \colhead{ROSAT\tablenotemark{a} }
  & \colhead{MF\tablenotemark{b} }
  & \colhead{Other IDs} }
\startdata
5 & H18(AGN) &       & \\
6 &          &       & 2MASS \\
11 &         & MF33  & \\
15 &         & MF34  & \\
37 & H22     &       & NGC5458, 2MASS, GSC \\
38 & H23     &       & $3''$ S of nucleus \\
40 & H23     &       & Nucleus, 2MASS, GSC \\
42 &         & MF46  & \\
43 &         & MF49  & \\
45 & H24     &       & \\
51 & H25     &       &  \\
52 &         & MF50  & \\
63 & H26     &       & \\
64 & H27     &       & \\
67 &         & MF54  & \\
70 & H29     &       & \\
76 & H30     &       & \\
81 &         & MF61  & \\
85 &         & MF65  & \\
95 &         &       & 2MASS \\
98 & H32(var) &      &  \\
99 & H33     &       & \\
103 & H35    &       & \\
104 & H36    & MF83  & \\
107 & H37    &       & NGC5461, 2MASS  \\
108 &        &       & 2MASS \\
110 & H40    &       & NGC5462, 2MASS  \\
\enddata
\tablenotetext{a}{{\em ROSAT} source ID from \citet{wan99} }
\tablenotetext{b}{SNR ID from \citet{mat97} }
\end{deluxetable}
\clearpage

%\end{document}

%% file: table3.tex
%\documentclass{aastex}
%\documentclass[preprint]{aastex}
%\begin{document}

\begin{deluxetable}{lllr}
\tablecaption{Energy Bands}
\tablewidth{0pt}

\tablehead{
\colhead{Band}  & \colhead{Low}   & \colhead{High}   & \colhead{Background}\\
\colhead{} & \colhead{keV}     & \colhead{keV}   & \colhead{cts pixel$^{-1}$}
}
\startdata

S1 & 0.125 & 0.8  & 0.034 \\
M1 & 0.8   & 1.3  & 0.007 \\
H1 & 1.3   & 8.0  & 0.032 \\
S2 & 0.125 & 0.5  & 0.023 \\
M2 & 0.5   & 2.0  & 0.025 \\
H2 & 2.0   & 8.0  & 0.025 \\

\enddata
\end{deluxetable}
\clearpage

%\end{document}

%% file: table4.tex
%\documentclass{aastex}
%\documentclass[preprint]{aastex}
%\begin{document}

\begin{deluxetable}{rrrrrcccccc}
\tablecaption{Xspec Model Fits to the 29 Brightest Sources}
\tablewidth{0pt}

\tablehead{
\colhead{}  & \colhead{}         & \colhead{}   & \colhead{}          &  \colhead{}                & \multicolumn{3}{c}{log($F_X$)}              & \multicolumn{3}{c}{log($L_X$)}            \\
\colhead{S} & \colhead{Pow}     & \colhead{BB} & \colhead{$N_H$}      & \colhead{$\chi^2_\nu$}    & \multicolumn{3}{c}{(ergs cm$^{-2}$ s$^{-1}$)}             & \multicolumn{3}{c}{(ergs s$^{-1}$)}   \\
\colhead{}  & \colhead{$\alpha$}& \colhead{kT}  & \colhead{($10^{21}$)} & \colhead{} & \colhead{S2} & \colhead{M2} & \colhead{H2} & \colhead{S2} & \colhead{M2} & \colhead{H2} 
}
\startdata

5 & 1.29 & \nodata & 0.65 & 1.46 & -14.92 & -13.88 & -13.38 & 37.56 & 37.99 & 38.41 \\
13 & \nodata & 0.10 & 0.10 & 2.06 & -14.50 & -14.87 & -20.06 & 37.48 & 36.95 & 31.74 \\
17 & 1.06 & \nodata & 3.66 & 1.00 & -17.06 & -14.67 & -13.84 & 36.83 & 37.40 & 37.96 \\
19 & 1.73 & \nodata & 0.62 & 1.40 & -15.35 & -14.52 & -14.28 & 37.19 & 37.35 & 37.51 \\
25 & 1.51 & \nodata & 5.97 & 0.75 & -17.19 & -14.25 & -13.53 & 37.70 & 37.99 & 38.29 \\
29 & 2.07 & \nodata & 4.56 & 1.13 & -16.65 & -14.44 & -14.06 & 37.84 & 37.79 & 37.75 \\
37 & \nodata & 0.13 & 1.92 & 1.74 & -15.40 & -14.63 & -17.75 & 37.68 & 37.58 & 34.07 \\
38 & 1.56 & \nodata & 0.35 & 1.30 & -14.97 & -14.25 & -13.95 & 37.32 & 37.58 & 37.85 \\
40 & 2.02 & \nodata & 0.34 & 1.30 & -14.72 & -14.22 & -14.19 & 37.63 & 37.62 & 37.61 \\
45 & \nodata & 0.06 & 0.78 & 1.34 & -14.34 & -14.89 & -23.68 & 38.50 & 37.14 & 28.12 \\
47 & \nodata & 0.15 & 4.96 & 1.39 & -16.35 & -14.55 & -16.67 & 37.95 & 38.04 & 35.20 \\
51 & 2.08 & \nodata & 0.84 & 1.13 & -15.28 & -14.48 & -14.42 & 37.47 & 37.42 & 37.38 \\
57 & 2.08 & \nodata & 10.14 & 0.76 & -18.46 & -14.47 & -13.83 & 38.10 & 38.06 & 38.01 \\
63 & 1.44 & \nodata & 0.39 & 0.58 & -14.95 & -14.14 & -13.76 & 37.36 & 37.70 & 38.03 \\
64 & 2.16 & \nodata & 3.11 & 1.57 & -16.26 & -14.58 & -14.34 & 37.66 & 37.56 & 37.46 \\
66 & 2.71 & \nodata & 19.56 & 1.24 & -21.25 & -14.69 & -14.00 & 38.76 & 38.33 & 37.90 \\
70 & 1.88 & \nodata & 2.02 & 1.54 & -15.11 & -13.71 & -13.42 & 38.24 & 38.31 & 38.38 \\
71 & 1.38 & \nodata & 0.47 & 1.59 & -15.21 & -14.32 & -13.90 & 37.15 & 37.53 & 37.90 \\
76 & 1.64 & \nodata & 4.60 & 1.11 & -16.13 & -13.70 & -13.12 & 38.26 & 38.48 & 38.69 \\
85 & 2.31 & \nodata & 0.54 & 1.14 & -15.21 & -14.69 & -14.80 & 37.37 & 37.18 & 36.99 \\
93 & 4.00 & \nodata & 7.57 & 1.38 & -16.30 & -13.99 & -14.26 & 40.01 & 38.80 & 37.60 \\
98 & \nodata & 0.14 & 0.53 & 4.11 & -13.17 & -12.78 & -15.77 & 39.11 & 39.12 & 36.03 \\
99 & \nodata & 0.08 & 0.63 & 1.81 & -14.44 & -14.73 & -21.32 & 38.11 & 37.24 & 30.48 \\
103 & 1.61 & \nodata & 1.04 & 0.75 & -15.36 & -14.26 & -13.91 & 37.42 & 37.65 & 37.89 \\
104 & 2.70 & \nodata & 1.23 & 1.19 & -14.54 & -13.80 & -14.04 & 38.60 & 38.18 & 37.76 \\
105 & 2.21 & \nodata & 9.74 & 0.73 & -18.31 & -14.51 & -13.94 & 38.15 & 38.02 & 37.90 \\
107 & \nodata & 0.13 & 3.75 & 1.12 & -15.41 & -14.13 & -16.89 & 38.46 & 38.38 & 34.96 \\
109 & \nodata & 0.20 & 0.20 & 0.98 & -15.25 & -14.68 & -16.56 & 36.75 & 37.15 & 35.24 \\
110 & 2.90 & \nodata & 2.27 & 1.39 & -14.79 & -13.71 & -13.93 & 38.96 & 38.42 & 37.87 \\
\enddata
\end{deluxetable}

\clearpage

%\end{document}

%% file: ms.bbl
\begin{thebibliography}{}
\bibitem[Arnaud(1996)]{arn96} Arnaud, K. A. 1996, in Astronomical Data
   Analysis Software and Systems V, ASP Conf Ser, Vol. 101, eds. Jacoby G.,
   and Barnes, J., 17
\bibitem[Asai et al.(1998)]{asa98} Asai, K., Dotani, T., Nagase, F, Ebisawa, K.,
   Mukai, K., Smale, A. P., and Kotani, T. 1998, \apjl, 503, L143
\bibitem[Bresolin et al.(1996)]{bre96} Bresolin, F., Kennicutt, R. C., Jr., 
   and Stetson, P. B. 1996, \aj, 112, 1009
\bibitem[de Vaucouleurs et al.(1991)]{dev91} de Vaucouleurs, G., de
   Vaucouleurs, A., Corwin, H., Jr., Buta, R., Paturel, G., and
   Fouque, P. 1991, Third Reference Catalogue of Bright Galaxies
   (Berlin:Springer)
\bibitem[de Vaucouleurs and Pence(1978)]{dev78} de Vaucouleurs, G. and
   Pence, W. D. 1978, \aj 83, 1163
\bibitem[Greiner et al.(1991)]{gre91} Greiner, J., Hasinger, G., and
   Kahabka, P. 1991, \aap, 246, L17
\bibitem[Freedman et al.(1994)]{fre94} Freedman, W. L., et al. 1994, \apj,
   427, 628
\bibitem[Hartmann and Burton(1997)]{har97} Hartmann, D. and Burton, W. B. 1997,
   Atlas of Galactic Neutral Hydrogen (Cambridge University Press)
\bibitem[Ho et al.(2001)]{ho01} Ho, L. C., et al. 2001, \apjl, 549, L51
\bibitem[Israel et al.(1975)]{isr75} Israel, F. P., Goss, W. M., 
   and Allen, R. J. 1975, \aap, 40, 421
\bibitem[Kamphuis(1993)]{kam93} Kamphuis, J. J. 1993, Ph.D. thesis, 
   Rijksuniversiteit, Groningen
\bibitem[Lampton et al.(1976)]{lam76} Lampton, M., Margon, B., and Bowyer, S.
   \apj, 208, 177
\bibitem[Long et al.(1981)]{lon81} Long, K. S., Helfand, D. J., and
   Grabelsky, D. A. 1981, \apj, 248, 925
\bibitem[Matonick and Fesen(1997)]{mat97} Matonick, D. M., and 
   Fesen, R. A. 1997, \apjs, 112, 49
\bibitem[McCammon and Sanders(1984)]{mcc84} McCammon, D. and 
   Sander, W. T. 1984, \apj, 287, 167
\bibitem[Mushotzky et al.(2000)]{mus00} Mushotzky, R. F., Cowie, L. L.,
   Barger, A. J., and Arnaud, K. A. 2000, \nat, 404, 459
\bibitem[Okamura et al.(1976)]{oka76} Okamura, S., Kanazawa, T., and
   Kodaira, K. 1976, \pasj, 28, 329
\bibitem[Shafter et al.(2000)]{sha00} Shafter, A. W., Ciardullo, R., 
   and Pritchet, C. 2000, \apj, 530, 193
\bibitem[Shanley et al.(1995)]{sha95} Shanley, L., \"Ogelman, H., 
   Gallagher, J. S., Orio, M., and Krautter, J. 1995, \apjl, 438, L95
\bibitem[Snowden et al.(2001)]{sno01} Snowden, S. L., Mukai, K.,
   Pence, W., and Kuntz, K. D. 2001, \aj, 121, 3000
\bibitem[Snowden and Pietsch(1995)]{sno95} Snowden, S. L., and Pietsch, W.
  1995, \apj, 452, 627
\bibitem[Stark et al.(1992)]{sta92} Stark, A. A., et al. 1992,
   \apjs, 79, 77
\bibitem[Stetson et al.(1998)]{ste98} Stetson, P. B., et al. 1998, \apj,
   290, 449
\bibitem[Supper et al.(1997)]{sup97} Supper, R., Hasinger, G., Pietsch, W.,
   Trumper, J., Jain, A., Magnler, E. A., Lewin, W. H. G., and van Paradijs, J.
   1997, \aap, 317, 328
\bibitem[Tennant et al.(2001)]{ten01} Tennant, A. F., Wu, K., Ghosh, K. K.,
   Kolodziejczak, J. J., and Swartz, D. A. 2001, \apjl, 549, L43
\bibitem[Thompson et al.(1998)]{tho98} Thompson, R. J. Jr, Shelton, R. G.,
   and Arning, C. A. 1998, \aj, 115, 2587
\bibitem[Trinchieri et al.(1990)]{tri90} Trinchieri, G., Fabbiano, G.,
   and Romaine, S. 1990, \apj, 356, 110
\bibitem[Trumpler et al.(1991)]{tru91} Trumpler, J., Hasinger, G., Aschenbach,
   B., et al., 1991, \nat, 349, 579
\bibitem[Van den Heuvel, et al.(1992)]{van92} Van den Heuvel, E. P. J., 
   Bhattacharya, D., Nomoto, K., Rappaport, S. A. 1992, \aap, 262, 97
\bibitem[Wang et al.(1999)]{wan99} Wang, Q. D., Immler, S., and
   Pietsch, W. 1999, \apj, 523, 121
\bibitem[Williams and Chu(1995)]{wil95} Williams, R., 
   and Chu, Y.-H. 1995, \apj, 439, 132

\end{thebibliography}
